\documentclass[fleqn,usenatbib]{mnras}

\usepackage{newtxtext,newtxmath}

\usepackage[T1]{fontenc}

\DeclareRobustCommand{\VAN}[3]{#2}
\let\VANthebibliography\thebibliography
\def\thebibliography{\DeclareRobustCommand{\VAN}[3]{##3}\VANthebibliography}

\usepackage{graphicx}	
\usepackage{amsmath}	
\usepackage{fix-cm}
\usepackage{subcaption}
\usepackage{float}
\usepackage{placeins}
\usepackage{MnSymbol}
\usepackage{enumerate}
\usepackage{threeparttable}
\usepackage{booktabs,caption}
\usepackage{xspace}
\usepackage{adjustbox}  

\newcommand\orcid[1]{\href{http://orcid.org/#1}{\adjustbox{trim={-.15\width} {0\height} {-.15\width} {0\height},clip}{\includegraphics[height=10pt]{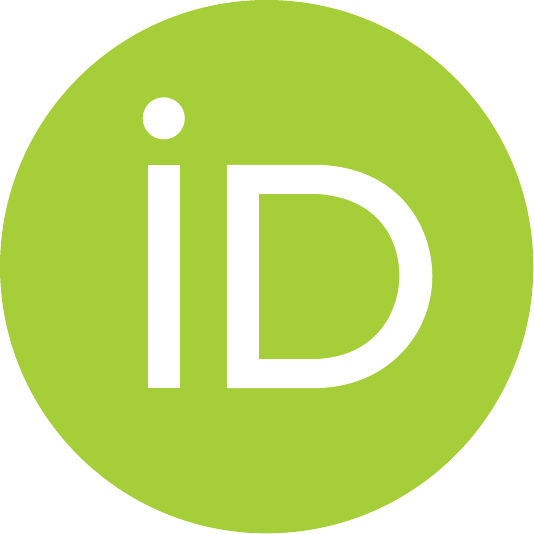}}}}

\usepackage{silence}
\title[Interaction driven star formation]{Mergers lighting the early Universe: modest star formation enhancement without evidence for AGN or Ly$\alpha$ triggering}

\author[Puskás et al.]{Dávid Puskás\orcid{0000-0001-8630-2031},$^{1,2}$\thanks{E-mail: dp670@cam.ac.uk}
Sandro Tacchella\orcid{0000-0002-8224-4505},$^{1,2}$
Charlotte Simmonds\orcid{0000-0003-4770-7516},$^{1,2,3}$
Gareth C. Jones\orcid{0000-0002-0267-9024},$^{1,2}$\newauthor
Ignas Juodžbalis\orcid{0009-0003-7423-8660},$^{1,2}$
Jan Scholtz\orcid{0000-0001-6010-6809}$^{1,2}$,
William M. Baker\orcid{0000-0003-0215-1104},$^{4}$
Andrew J.\ Bunker\orcid{0000-0002-8651-9879},$^{5}$\newauthor
Stefano Carniani\orcid{0000-0002-6719-380X},$^{6}$
Emma Curtis-Lake\orcid{0000-0002-9551-0534},$^{7}$
Qiao Duan\orcid{0009-0009-8105-4564},$^{1,2}$
Daniel J.\ Eisenstein\orcid{0000-0002-2929-3121},$^{8}$\newauthor
Kevin Hainline\orcid{0000-0003-4565-8239},$^{9}$
Benjamin D.\ Johnson\orcid{0000-0002-9280-7594},$^{8}$
Roberto Maiolino\orcid{0000-0002-4985-3819},$^{1,2, 10}$
Marcia Rieke\orcid{0000-0002-7893-6170},$^{9}$\newauthor
Brant Robertson\orcid{0000-0002-4271-0364},$^{11}$
Christina C. Williams\orcid{0000-0003-2919-7495},$^{12}$
and Joris Witstok\orcid{0000-0002-7595-121X}$^{13,14}$
\\
\\
$^{1}$Cavendish Laboratory, University of Cambridge, 19 JJ Thomson Avenue, Cambridge, CB3 OHE, UK\\
$^{2}$Kavli Institute for Cosmology, Madingley Road, Cambridge, CB3 0HA, UK\\
$^{3}$Departamento de Astronomía, Universidad de Chile, Camino El Observatorio 1515, Las Condes, 7591245 Santiago, Chile\\
$^{4}$DARK, Niels Bohr Institute, University of Copenhagen, Jagtvej 128, DK-2200 Copenhagen, Denmark\\
$^{5}$Department of Physics, University of Oxford, Denys Wilkinson Building, Keble Road, Oxford OX1 3RH, UK\\
$^{6}$Scuola Normale Superiore, Piazza dei Cavalieri 7, I-56126 Pisa, Italy\\
$^{7}$Centre for Astrophysics Research, Department of Physics, Astronomy and Mathematics, University of Hertfordshire, Hatfield AL10 9AB, UK\\
$^{8}$Center for Astrophysics $|$ Harvard \& Smithsonian, 60 Garden St., Cambridge MA 02138 USA\\
$^{9}$Steward Observatory, University of Arizona, 933 N. Cherry Avenue, Tucson, AZ 85721, USA\\
$^{10}$Department of Physics and Astronomy, University College London, Gower Street, London WC1E 6BT, UK\\
$^{11}$Department of Astronomy and Astrophysics University of California, Santa Cruz, 1156 High Street, Santa Cruz CA 96054, USA\\
$^{12}$NSF National Optical-Infrared Astronomy Research Laboratory, 950 North Cherry Avenue, Tucson, AZ 85719, USA\\
$^{13}$Cosmic Dawn Center (DAWN), Rådmandsgade 64, 2200 København, Copenhagen, Denmark\\
$^{14}$Niels Bohr Institute, University of Copenhagen, Jagtvej 128, DK-2200, Copenhagen, Denmark} 

\pubyear{\the\year{}}

\begin{document}
\label{firstpage}
\pagerange{\pageref{firstpage}--\pageref{lastpage}}
\maketitle

\begin{abstract}
Galaxy mergers and interactions are often invoked to explain enhanced star formation, black hole growth, and mass build-up of galaxies at later cosmic times, but their effect is poorly understood at high redshift ($z>2$). We use JADES data to analyse a mass-complete sample of 1968 galaxies at $z=3-9$ with ${\rm log}(M_\star/{\rm M_\odot}) = [8, 10]$, identifying major merger pairs (projected separation of $5-100$ pkpc, mass ratio $\geq 1/4$) using a probabilistic method. To look for signatures of enhancement in multiple physical properties, we carefully build a control sample of non-pairs that are simultaneously matched in redshift, stellar mass, isolation, and environment to the pair sample. We find a moderate enhancement in specific star-formation rate (sSFR) of a factor of $1.10\pm 0.05$ at separations $\lesssim 20$~kpc in pairs compared to their matched controls. We find that the interaction-related sSFR enhancement is most apparent at longer SFR averaging time-scales (50–100 Myr), whereas the enhancement is not detected on 5–10 Myr time-scales, where bursty star formation might be driven by internal processes. By averaging star formation histories, we find two distinct populations: one with monotonically rising SFRs and another with an earlier peak in SFR, potentially corresponding to different interaction stages. Finally, we find no significant excess of AGN in pairs, consistent with galaxy interactions not strongly triggering black hole activity at the separations probed ($>5$~kpc). Similarly, we also do not detect an excess in the fraction of Lyman-$\alpha$ emitters in pairs, suggesting that at the probed separations, galaxy interactions are not efficient at enhancing Lyman-$\alpha$ photon production and escape, which may only become important at the smallest physical scales.
\end{abstract}




\section{Introduction}

Galaxy mergers and interactions play a key role in shaping the evolutionary path of individual galaxies. In the Lambda Cold Dark Matter ($\Lambda$CDM) cosmological framework, dark matter haloes merge under the hierarchical growth model, resulting in the tidal disruption and eventual merger of the galaxies embedded within them \citep{Sommerville_2015}. It has been well established that merging systems can undergo profound structural transformations \citep{Toomre_1972}, elevated star formation rates (SFRs) \citep{Barton_2000, Ellison_2008, Scudder_2012, Bickley_2022}, and enhanced active galactic nuclei (AGN) activity \citep{Ellison_2011, Ellison_2013, Ellison_2019}. Most of these results come from large low-redshift samples. At high redshift, the abundance of merger events has only recently begun to be constrained at $z>6$ \citep{Duncan_2019, Duan_2025, Puskas_2025}, and the effects of these mergers on galaxy properties remain poorly understood.

Theory and simulations predict that the gravitational interaction during close galaxy-galaxy encounters and mergers can induce non-axisymmetric features (e.g., bars) and strong torques that channel low-metallicity gas to galactic centres, fuelling starbursts and potentially triggering AGN \citep{Mihos_1996, Springel_2005, Cox_2008, Scudder_2012, Torrey_2012, Hopkins_2013, Patton_2013, Moreno_2015, Patton_2020, Byrne-Mamahit_2024}. For similar-mass mergers, gas inflows can drive short, intense bursts (enhancements of order a factor of $\sim$2), with the strength depending primarily on internal structure and mass ratio rather than orbital parameters \citep{Mihos_1996, Cox_2008}. Gas-rich mergers may then fuel black hole (BH) accretion and feedback, expelling (or using up) gas near coalescence and contributing to quenching \citep{Springel_2005}. Many simulation-based works describe moderate SFR enhancement for several hundred Myr after first pericentre and a stronger burst shortly before coalescence \citep[e.g.,][]{Mihos_1996, Cox_2008, Scudder_2012}, with the post-coalescence starburst lasting up to $\sim\!0.5$~Gyr \citep{Hani_2020, Ferreira_2025} and potentially producing quenched elliptical remnants \citep{Quai_2021, Ellison_2022}.

Observational studies of galaxy mergers typically focus on two aspects: (i) the frequency of mergers (pair fractions and merger rates) as a function of redshift, and (ii) the impact on galaxy properties. The merger rate has been well constrained at lower redshifts ($z<3$) \citep[see, e.g.,][]{Man_2016, Mundy_2017, Mantha_2018, Conselice_2022}, with several works extending to higher redshifts $z \sim 6-7$ \citep[e.g.,][]{Duncan_2019, Ventou_2017}. Only a few recent works \citep{Dalmasso_2024, Duan_2025, Puskas_2025} probed the highest redshifts beyond $z>6$ thanks to the wealth of high-resolution photometry and spectroscopy from the \textit{James Webb Space Telescope}(\textit{JWST}). These works found an initially steeply increasing, then a flattening merger rate with redshift, stabilising in the range of $2-10$ major mergers/galaxy/Gyr at $z>6$ \citep{Puskas_2025}. To infer the merger rate, two widely used methods exist for selecting galaxies that are about to merge: the \textit{close-pair method} that involves counting galaxies that are in close pairs on the projected plane of the sky \citep[e.g.,][]{Mundy_2017, Duncan_2019, Duan_2025, Puskas_2025}, and the \textit{morphological method} that involves finding systems showing disturbed morphologies that are at the late stages or have recently completed merging \citep[e.g.,][]{Conselice_2008, Rose_2023, Dalmasso_2024}. Here we focus on the effects of close encounters on internal galaxy properties, rather than the merger rate itself, which has been extensively discussed in \citet{Puskas_2025}.

The properties of galaxy mergers and their influence on SFRs have been studied in depth using large spectroscopic samples at low redshift ($z < 1$), such as the CfA2 \citep{Barton_2000} and 2dF \citep{Lambas_2003} surveys. A wealth of studies using data from the Sloan Digital Sky Survey (SDSS; latest data release DR19 \citealt{SDSS_2026}) have shown that galaxies in close pairs with projected separations lower than 80 kpc exhibit enhanced SFRs (on average 60 per cent higher) in comparison to a carefully matched control sample of isolated galaxies \citep{Nikolic_2004, Ellison_2008, Scudder_2012, Ellison_2013, Patton_2013, Patton_2016}. \citet{Patton_2013} have found moderate SFR enhancements at separations $20-100$~kpc, and smaller enhancements even up to $\!\sim\!150$~kpc. This was obtained through careful selection and matching of control samples to compare galaxy close pairs with galaxies that have very similar physical properties but are isolated.

However, at higher redshift ($z>1$), there is less agreement between different theoretical and observational studies regarding the enhancement (if any) or even the decrease of SFR between merger and non-merger samples. Simulations by \citet{Fensch_2017} demonstrate that high-redshift (up to $z \sim 4$), gas-rich galaxy mergers are substantially less efficient at enhancing their SFRs than their low-redshift, gas-poor counterparts, producing starbursts that are approximately ten times weaker and shorter in duration. Observational studies also find different trends at higher redshifts. For example, at $0.5 < z < 3$, \citet{Shah_2022} have found a somewhat weaker, but still detectable enhancement (factor of $\!\sim\!1.23^{+0.08}_{-0.09}$) at the closest separations of $\lesssim 25$~kpc. Probing to the highest redshift to date using novel \textit{JWST} data, \citet{Duan_2026} have only detected weak enhancements of SFRs at the smallest separations of $r_{\rm p}< 20$~kpc of $0.25 \pm 0.10$ dex and $0.26 \pm 0.11$ dex above the non-merger medians for the redshift ranges of $4.5< z <6.5$ and $6.5 < z < 8.5$, respectively. On the other hand, \citet{Silva_2018} find no significant difference in SFR between merging and non-merging galaxies at $r_{\rm p} < 15$ and $0.5 < z < 2.5$, with only a small fraction (12 per cent) of massive systems showing starburst activity. Using a machine learning approach on over 200,000 galaxies, \citet{Pearson_2019} found that while mergers typically have little to no effect on the average SFR (typical change in SFR is less than 0.1 dex in either direction), the fraction of galaxies identified as mergers steadily increases in systems with the most highly-enhanced, starbursting activity. Therefore, the exact effect of mergers and its magnitude on the SFR of the constituent galaxies remain uncertain at redshift $z>3$.

Furthermore, mergers have been found to efficiently trigger AGN activity and drive BH accretion \citep[e.g.][]{Ellison_2011, Ellison_2013, Byrne-Mamahit_2024}. Simulations \citep[e.g.,][]{DiMatteo_2005} show that gravitational torques during an interaction can funnel vast amounts of gas to the galactic centre, igniting both a starburst and a luminous quasar. Observational evidence provides a more nuanced picture. Studies of large galaxy samples find a modest enhancement \citep[e.g., a factor of 2-3 found by][]{Ellison_2011, Ellison_2013} in the fraction of AGN in close galaxy pairs (separations $< 20-30$~kpc), indicating that the final stages of mergers are most impactful. This link appears strongest for the most powerful and obscured AGN. For example, work by \citet{Koss_2010} and \citet{Satyapal_2014} found that luminous, hard X-ray or IR-selected AGN are significantly more likely to reside in disturbed, merging systems than their inactive counterparts. Other studies confirm that post-mergers host an excess of AGN compared to isolated galaxies \citep[e.g.,][]{Bickley_2023, Byrne-Mamahit_2024}. However, since the overall fraction of AGN in these merging populations remains low, it is unclear if mergers guarantee AGN triggering. Furthermore, this connection appears to weaken with lookback time, as studies of high-z interacting galaxies often find no statistically significant AGN excess \citep[e.g.][]{Shah_2020, Silva_2021, Byrne-Mamahit_2024, Dougherty_2024}.

Moreover, Lyman-$\alpha$ emitters (LAEs) at high-z are often found in ionized bubbles already present in the Epoch of Reionization (EoR), embedded in large overdensities \citep[e.g.,][]{Larson_2022, Leonova_2022, Whitler_2024}. The presence of such overdensities \citep[e.g.,][]{Helton_2024b} can boost galaxy interactions and mergers, as well as potentially lead to subsequent quenching \citep[e.g.,][]{Carnall_2024, Baker_2026, deGraaff_2025, Cochrane_2025}. The resulting merger-induced starbursts and potential AGN activity can provide the powerful ionizing radiation required to illuminate the surrounding reservoirs of hydrogen gas and may create disturbed, clumpy ISM/CGM morphologies, drive strong outflows, and carve low-density channels that allow efficient Ly-$\alpha$ photon escape. Recent observations with \textit{JWST} unveiled a large population of LAEs at redshift $z \sim 5-8$ deep into the EoR \citep[e.g.,][]{Saxena_2023, Witstok_2024, Jones_2025a}, which show morphological disturbances and have resolved close companions \citep[e.g.,][]{Witten_2024, Ren_2026}. It is yet to be understood what fraction of these LAEs are found to be in close pairs and if mergers on a population level effectively trigger them.

The aim of this study is to investigate whether the presence of a close companion has any influence on the physical properties of its host galaxy, such as star formation or mass growth. Furthermore, we assess if this pre-merger phase of close galaxy pairs triggers any AGN activity and Ly$\alpha$ emission that would deepen our understanding of the EoR. We aim to compare our mass-complete sample of major merger pairs ($r_{\rm p}=5$–100\,kpc, mass ratio $\geq1/4$, $\log_{10}(M_\star/{\rm M_\odot})=8$–10) to a robust control set \emph{simultaneously matched} in redshift, stellar mass, local density, and isolation, enabling us to isolate the incremental effect of the nearest neighbour from environmental trends. This has been done by several studies for the local Universe, however, in this work, we extend the analysis for the $3 \leq z \leq 9$ redshift range using deep \textit{JWST} data.

The paper is structured as follows. Section~\ref{sec:sample selection} provides a brief overview of the catalogues and data products used in this study. We then describe galaxy pair selection and its environmental characterisation, and detail the control matching process. Section~\ref{sec:Enhanced Star Formation in Galaxy Pairs} describes the influence of the companion on the observed physical properties of the host galaxies compared to the control sample, such as star formation rate and star formation history, dependence on stellar mass and redshift, as well as environmental dependence. In Section~\ref{sec:merger driven AGN and LAE} we discuss the fraction of AGN and Ly$\alpha$ emitters found in pairs compared to the overall spectroscopic pair fraction. Section~\ref{sec:discussion} discusses our findings and their implications in a broader cosmological context, as well as some caveats of our study and the effect of parameter choices. Finally, Section~\ref{sec:conclusion} summarises our work and highlights the main findings.

Throughout this paper, we adopt the AB magnitude system \citep{Oke_1983}. We use a standard cosmology with $\Omega_{m} = 0.310$,  $\Omega_{\Lambda} = 0.689$, and $H_0 = 67.66 \mathrm{\ km \ s^{-1} \ Mpc^{-1}}$ \citep{Planck_2020}. Throughout our analysis, we use the \texttt{Astropy} python package \citep{astropy:2022}, and its subpackage \texttt{astropy.cosmology}, where we assume a flat $\Lambda$CDM cosmology with parameters from \citet{Planck_2020}.

\section{Sample and methodology}
\label{sec:sample selection}

In Section~\ref{sec:data}, we briefly summarise the data sets used from the JADES survey, and the photometric and spectroscopic redshift catalogues. Section~\ref{sec:SED modelling} presents the spectral energy distribution (SED) modelling and the assumptions that go into it. The probabilistic galaxy pair selection process for finding major mergers is presented in Section~\ref{sec:galaxy pair selection}. In Section~\ref{sec:environmental characteristaion}, we characterise the environment of the galaxies in our target sample, which contains both pairs and control candidates. Finally, in Section~\ref{sec:control sample}, we present the construction of a robust control sample by simultaneously matching pairs in both stellar mass and redshift space, as well as local density and isolation, to compare the physical properties of pairs against those of the controls.

\subsection{Data}
\label{sec:data}

This study is based on the rich photometric and spectroscopic data from the \textit{JWST} Advanced Deep Extragalactic Survey \citep[JADES, ][]{Eisenstein_2026, Rieke_2023, Eisenstein_2025, Bunker_2024, D'Eugenio_2025}. We divide the two constituent JADES fields, GOODS-South and GOODS-North, into four different sub-tiers based on the area and exposure time. The deep tiers are defined by exposure-time thresholds in the F444W band of $T_{\rm exp}^{\rm GS}=32.5$ ks and $T_{\rm exp}^{\rm GN}=9.5$ ks for GOODS-South and GOODS-North, respectively, with the remaining regions forming the corresponding medium tiers. For a detailed description of the various data sets used, we refer the reader to Section 2 of \citet{Puskas_2025}. We briefly summarise here the photometric and spectroscopic redshifts used for this study, and the SED modelling in the following section.

The photometric catalogues include JADES NIRCam imaging in F090W, F115W, F150W, F162M, F200W, F250M, F277W, F300M, F335M, F356W, F410M, and F444W bands. This is supplemented, where available, by the JEMS medium-band observations in F182M, F210M, F430M, F460M, and F480M \citep{Williams_2023}, as well as ancillary \textit{HST} photometry in the ACS F435W, F606W, F775W, F814W, and F850LP bands and the infrared F105W, F125W, F140W, and F160W bands. We use Kron apertures measured on images convolved to a common resolution (\texttt{KRON\_CONV}) in F444W to obtain consistent photometry across the different bands.

We obtain photometric redshifts for all galaxies using the template-fitting code \texttt{EAZY} \citep{Brammer_2008}. A detailed description of the templates used and assumptions made for our fits can be found in section 3.1 of \citet{Hailine_2024}. \texttt{EAZY} searches for the best-fitting linear combination of galaxy templates over a redshift grid of $z=0.01-22$ with a step size of $\Delta z=0.01$. We use modified versions of the original seven \texttt{EAZY} templates that include emission-line contributions, supplemented by a dusty and a high-equivalent-width template, as well as seven JADES templates optimised using mock observations from the JAGUAR simulations \citep{Williams_2018}. These latter templates were constructed using the Flexible Stellar Population Synthesis (\texttt{FSPS}; \citealt{Conroy_2010}) framework and extend the colour space covered by the templates towards, in particular, blue UV-bright and red dusty galaxies (see \citealt{Hailine_2024} for details. Our adopted redshift is taken from the minimum in the $\chi^2$ of the fit. For each source, we obtain the photometric redshift posterior distribution by assuming a constant prior for the redshift, and calculate $P(z) = \textrm{exp}[-\chi^2(z) / 2]$ with a normalisation of $\int P(z) dz = 1.0$. To select a robust sample of galaxies with accurate photometric redshifts, we use the \textit{odds quality parameter} $\mathcal{O}$ as a proxy for the reliability of the photometric redshift fit, which is defined as
\begin{equation}
\mathcal{O} = \int^{+K (1 + z_a)}_{-K(1+z_a)} P(z-z_a) dz,
\end{equation}
with $K=0.03$ chosen for our analysis. In our sample selection, we require that all objects must have an odds parameter $\mathcal{O} \geq 0.3$.

Furthermore, we assemble a catalogue of all available spectroscopic redshifts, which consists of 5382 sources in the GOODS-South and 2591 in the GOODS-North field, all of which fall within the best category in quality and coordinate match. This catalogue combines spectroscopic measurements from JADES NIRSpec \citep{Bunker_2024, D'Eugenio_2025}, FRESCO and CONGRESS redshifts (F. Sun, private communication; see also \citealt{Meyer_2024} for H$\alpha$, \citealt{Covelo-Paz_2025} for H$\beta$+[O {\sc III}], and the survey paper by \citealt{Oesch_2023}), MUSE-Wide DR1 \citep{Urrutia_2019}, MUSE DR2 \citep{Bacon_2023}, ASPECS (ALMA SPECtroscopic Survey in the UDF) Large Program \citep{Decarli_2019}, and legacy spectroscopic observations in the CANDELS fields (N. Hathi, private communication). The quality flags from the different surveys are converted into a common classification, and we construct a catalogue of the best available redshifts in which a secure spectroscopic redshift is adopted in preference to the photometric estimate whenever available. If multiple highest-quality spectroscopic measurements are available and agree within $\Delta z_{\rm spec}<0.1$, we use their mean, while the small number of cases with discrepant secure measurements are excluded. For a detailed description of this catalogue, see Section 2 of \citet{Puskas_2025} and references therein.

\subsection{SED modelling and star formation rates}
\label{sec:SED modelling}

Stellar population properties were derived using the SED modelling code \texttt{Prospector} \citep{Johnson_2019,Johnson_2021}. We refer the reader to \citet{Simmonds_2024} and \citet{Simmonds_2025} for a detailed description of the fitting methodology, priors, and assumptions, and briefly summarise the aspects most relevant for this work.

The full available \textit{HST} and \textit{JWST} photometry described in Section~\ref{sec:data} is used as input to \texttt{Prospector}. The continuum and emission properties of the model SEDs are generated using the \texttt{FSPS} framework. We assume a Chabrier initial mass function \citep{Chabrier_2003} with lower and upper mass cut-offs of $0.1$ and $100\,{\rm M_\odot}$, respectively, and a top-hat prior on stellar mass over $\log_{10}(M_\star/{\rm M_\odot})=[6,12]$. The stellar metallicity is allowed to vary between $0.01$ and $1\,{\rm Z_\odot}$, and nebular line and continuum emission are included, with a free ionization parameter limited to $\log_{10}(U)\leq-1.0$. We adopt a two-component dust attenuation model \citep{Charlot_2000, Conroy_2009} that accounts for the different attenuation experienced by young stars ($<10$ Myr) and nebular emission, with a variable dust attenuation curve \citep{Kriek_2013}. A flexible model for intergalactic-medium absorption based on a scaling of the \citet{Madau_1995} prescription is also included. The \texttt{EAZY} photometric redshift constraints are used as priors in the SED modelling, while the redshift is fixed to the spectroscopic value when a secure measurement is available.

Of particular importance is the use of a non-parametric star-formation history (SFH) with the continuity prior of \citet{Leja_2019}, in which the SFH is described by eight time bins. The first bin extends to a lookback time of 5 Myr (relative to the redshift of the galaxy), and the last bin is fixed at the lookback time corresponding to $z = 20$, with the remaining time bins divided into equal intervals of $\log_{10}(t_{\rm lookback})$. The ratios and amplitudes between adjacent bins are allowed to vary according to a Student's $t$-distribution with a scale width of 0.3, which permits -- but does not enforce -- a modest level of burstiness when supported by the data. From the reconstructed SFHs, we compute time-averaged SFRs on different timescales $t_{\rm avg}$ according to
\begin{equation}
\mathrm{SFR}_{t_{\mathrm{avg}}} =
\frac{1}{t_{\mathrm{avg}}}
\int_{0}^{t_{\mathrm{avg}}}
\mathrm{SFR}(t^\prime)\,dt^\prime,
\label{eq:time-averaged-sfr}
\end{equation}
where $t^\prime$ is the lookback time. For this study, we primarily adopt $\mathrm{SFR}_{100}$ (SFR averaged over 100 Myr) as our fiducial measure of recent star formation, while also exploring shorter ($5-50$ Myr) and longer (2000 Myr) averaging windows to assess the dependence of our results on timescale.

\subsection{Galaxy pair selection}
\label{sec:galaxy pair selection}

We aim to investigate the influence of close companions on each galaxy in our selected target sample, which we define below, to statistically measure the effect of mergers on the physical properties of the host galaxies. Our selection of galaxy pairs strongly relies on the methodology outlined in Section 3 of \citet{Puskas_2025}. However, in this work we do not exclusively study \textit{close pairs}, defined as galaxy pairs having a projected physical separation at $5 \ {\rm kpc} \leq r_{\rm p} \leq 30 \ {\rm kpc}$, but we extend our search to larger separations (up to $r_{\rm max} = 100 \ {\rm kpc}$) to select \textit{wide pairs}. We simply refer to these selected systems as \textit{galaxy pairs} for the remainder of this paper. We note here that we define the projected physical separation as $r_{\rm p}=\theta \times d_{A}(z)$, where $\theta$ is the measured angular separation between the pairs and $d_{A}(z)$ is the angular diameter distance. In the following, we summarise the selection criteria from \citet{Puskas_2025}, which is further based on the probabilistic close-pair selection method developed by \citet{Lopez-Sanjuan_2015} and \citet{Mundy_2017}. The strength of this method lies in using the full photometric redshift probability distribution function to propagate the associated uncertainties, as opposed to only using the peaks of these distributions, as well as incorporating all available spectroscopic redshifts.

Briefly, for each potential projected pair we first quantify its overlap in redshift space using the individual photometric redshift posterior distributions $P_1(z)$ and $P_2(z)$, defining the redshift probability function as $\mathcal{Z}(z)=2 \times P_1(z) \times P_2(z)/[P_1(z)+P_2(z)]$. When a secure spectroscopic redshift is available, it is incorporated into the same probabilistic framework by replacing the photometric posterior with a narrow Gaussian centred on $z_{\rm spec}$ with $\sigma_z=0.01$. The redshift probability is then combined with an angular-separation mask, $\mathcal{M}^{\theta}(z)$, which selects only redshifts at which the observed angular separation corresponds to $r_{\rm min}\leq r_{\rm p}\leq r_{\rm max}$, and a stellar-mass selection mask, $\mathcal{M}^{\rm pair}(z)$, which imposes the major merger mass-ratio and stellar mass completeness criteria. The resulting pair probability function is therefore defined as $\mathrm{PPF}(z)=\mathcal{Z}(z)\times\mathcal{M}^{\theta}(z)\times\mathcal{M}^{\rm pair}(z)$, and its integral over redshift quantifies the probability contribution of the two galaxies to a galaxy major merger pair.

We are interested in the influence of close companions exerted on the central or host galaxy, which we refer to as the \textit{primary galaxy}. This is defined as the most massive galaxy in the system, and it is selected from a primary sample, which is described in a later section. In this study, we are only interested in major merger systems, which are defined as having a stellar mass ratio between the secondary and the primary galaxy of $\mu = M_2 / M_1 \geq 0.25$ \citep[e.g.,][]{Lotz_2011}. For our major merger pair catalogue, we apply the selection criteria described below.

\begin{enumerate}[1.]

\item The primary galaxy must be in the initial primary sample catalogue (detailed later in this section), being in the redshift range of $3 \leq z \leq 9$ and stellar mass range of $\log_{10}(M_\star /{\rm M_\odot}) = [8, 10]$.

\item The projected physical separation to the closest companion must be within $r_{\mathrm{min}} \leq r_{\rm p} \leq r_{\mathrm{max}}$, with our choice of $r_{\mathrm{min}}=5 \ {\rm kpc}$, to avoid confusing star-forming clumps within the same galaxy with mergers and ensure clearly deblended sources, and $r_{\mathrm{max}} = 100 \ {\rm kpc}$, to assess the influence of companions out to wide enough separations. We convert this projected physical separation to an observable angular separation using the angular diameter distance calculated at the peak redshift of the central galaxy.

\item To ensure close proximity in the radial direction dictated by the redshift, we require the integral of the \textit{pair probability function} described above \citep[see also Section 3 of][]{Puskas_2025} to be $\int_{3}^{9} \mathrm{PPF}(z) dz > 0.7$. This quantity incorporates the full photometric redshift posterior distributions from \texttt{EAZY} and gives the probability of a system of two galaxies being in a major merger pair.

\item To find major merger pair systems, we require a stellar mass ratio above $\mu = M_2 / M_1 \geq 0.25$.

\end{enumerate}

We select galaxies from an initial sample defined in detail in \citet{Puskas_2025}, resulting in a sample of 2095 galaxies in pairs, focusing on the properties of the primary (more massive) galaxies in these pairs. In summary, we perform the selection separately in four distinct JADES sub-fields, namely the GOODS-South (GS) and GOODS-North (GN) Deep and Medium tiers. We require a signal-to-noise ratio of $\mathrm{SNR} \geq 3$ in the \texttt{F444W\_KRON\_CONV} photometry (Kron-aperture placed on images that have been convolved to the same resolution). We select galaxies in the redshift range of $3 \leq z \leq 9$, where most objects have photometric redshifts estimated by \texttt{EAZY}, and additionally, we use all available spectroscopic redshifts (detailed in \citealt{Puskas_2025}). In the case of photometric redshifts, we require an odds quality parameter $\mathcal{O} \geq \mathbf{0.3}$ for robust $z_{\rm phot}$ measurements. Furthermore, we define our initial primary sample to have stellar masses estimated by \texttt{Prospector} in the range of $\log_{10}(M_\star / \mathrm{M_\odot}) = [8, 10]$. We adopt the upper limit of $\log_{10}(M_\star/{\rm M_\odot})=10$ to remain consistent with the close-pair analysis of \citet{Puskas_2025}, while galaxies above this mass become increasingly rare towards higher redshifts, resulting in limited sample sizes (see Fig.~\ref{fig:mass_completeness}). The secondary sample of galaxies, which contains all potential companions corresponding to the primary sample, extends to lower masses, as the stellar mass ratio is fixed for major mergers at $\mu \geq 0.25$. Both the primary and the secondary sample have to be above the mass completeness limit $M_\star^{\mathrm{comp}}(z)$, which is determined separately for the four tiers individually based on their 5$\sigma$ photometric depths (in the F444W band) using the method by \citet{Pozzetti_2010}. The adopted lower stellar-mass limit of the primary sample, $\log_{10}(M_\star/\mathrm{M_\odot})=8$, lies above the estimated 90 per cent stellar mass completeness limit throughout the full redshift range considered. We present the stellar mass distribution and the completeness calculation in Appendix~\ref{app:mass_completeness}.

\subsection{Environmental characterisation}
\label{sec:environmental characteristaion}

\begin{figure}
 \includegraphics[width=\columnwidth]{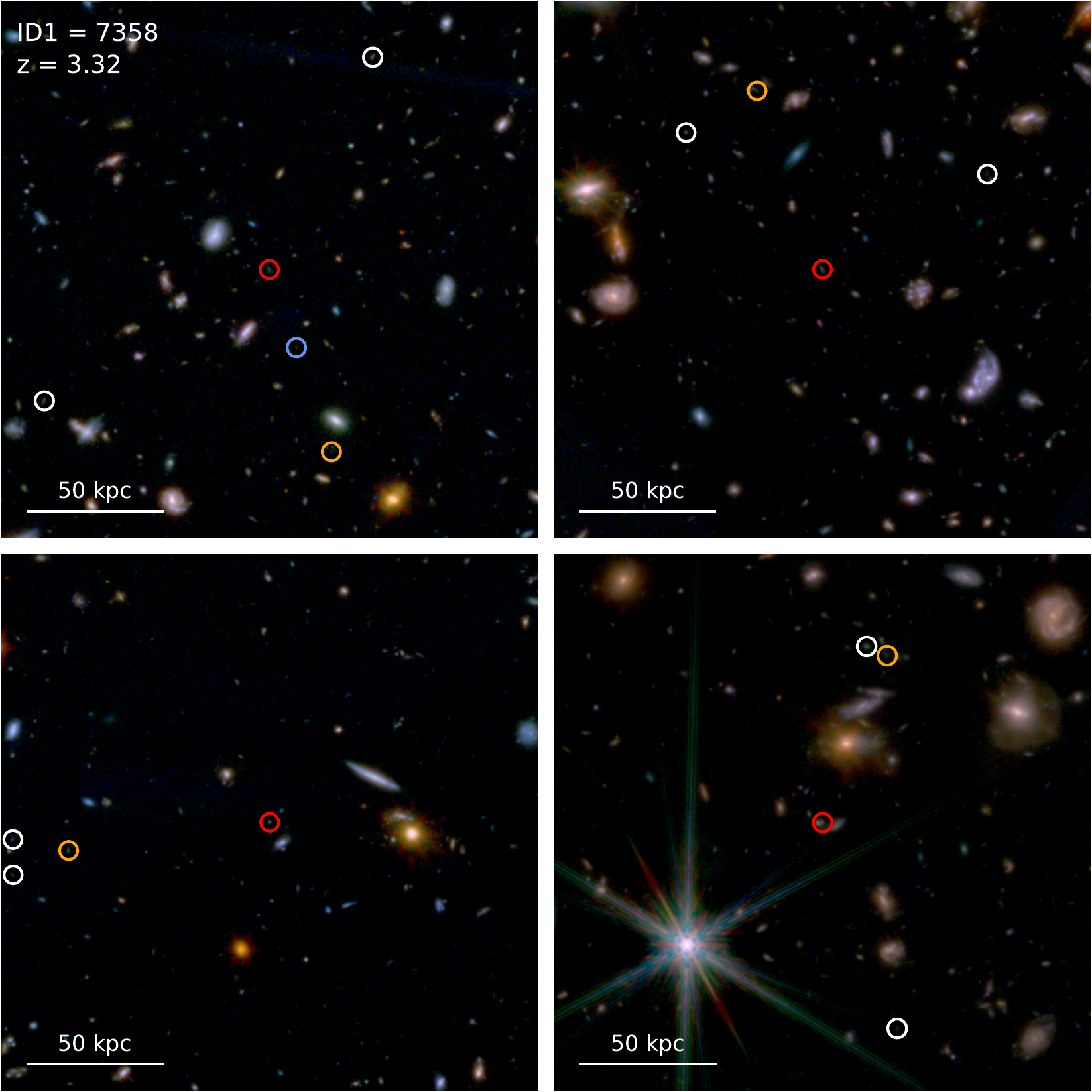}
 \caption{An example galaxy pair system with its environment (top left) 
 and three closely matched control systems. The primary galaxy is at the centre of each cutout, indicated by a red circle. The secondary galaxy in the pair system is indicated by a blue circle (only in the top-left cutout). The second closest companion is indicated by an orange circle, while for the three control systems the orange circles indicate the closest companions at comparable distances to the pair system's second closest neighbour. All other galaxies counted in the local density measurement $N_{\rm env}$ are indicated by white circles. The cutouts are limited to $200 \times 200$ kpc for display purposes, while the environment is analysed within a projected physical radius of 1 Mpc. The focus is on the relative projected positions of the galaxies rather than their individual appearance.}
 \label{fig:pairs_and_controls}
\end{figure}

The aim of this study is to investigate the influence of the closest companion on the physical properties of the primary galaxy. However, to properly assess this influence, one must also consider the surrounding environment of the pair. This has been done in previous studies of the local Universe, typically at $z < 0.2$, using various metrics to characterise the environment \citep[e.g.,][]{Ellison_2010, Scudder_2012, Patton_2013, Patton_2016, Garduno_2021}, with only a few works extending up to $z\sim3{}$ \citep[e.g.,][]{Shah_2022}.

For the purpose of this work, we apply the same metrics to characterise the galaxy environment as in \citet{Patton_2013} and \citet{Patton_2016}, namely, \textit{local density} and \textit{isolation}. These properties are measured in various ways by different studies, for example \citet{Ellison_2010} uses projected galaxy density averaged from the distances to the fourth and fifth nearest neighbours within 1000 ${\rm km \ s^{-1}}$, and \citet{Scudder_2012} measures the density via the distance to the fifth nearest neighbour.

To measure the \textit{local density}, we use a simplified metric $N_{\rm env}$, which gives the total number of companions around a galaxy within a projected physical separation of 1 Mpc. We choose this separation limit as it is an order of magnitude larger than the limit for selecting galaxy pairs. In our fiducial analysis, galaxies counted in $N_{\rm env}$ have to both satisfy the mass ratio criterion of $\mu \geq 0.25$, as well as $\int_{3}^{9} \mathrm{PPF}(z) dz > 0.1$. This latter limit on the pair probability function integral is lower than in the case of the closest companion ($\geq 0.7$) as we want to include all possible surrounding galaxies in the neighbourhood (even if they have a low probability), and the radial proximity criterion also becomes looser at separations comparable to $\sim 1$ Mpc. We note that, because of this deliberately less restrictive probability threshold, individual galaxies included in the environmental measurements are not necessarily secure physical companions. Instead, $N_{\rm env}$ and $r_2$ should be interpreted as statistical measures of the surrounding galaxy environment. For paired galaxies, the selected closest companion is included in the calculation of $N_{\rm env}$.

The \textit{isolation} of a galaxy pair in terms of environment can be measured by the projected distance to the second closest companion, denoted by $r_2$. The value of this parameter provides insight into the environment of a galaxy pair: high $r_2$ values indicate relative isolation, intermediate $r_2$values correspond to more typical field environments, and low $r_2$ values are characteristic of more densely populated regions, such as galaxy groups or overdensities. The second-closest neighbour must satisfy the same criteria as galaxies considered in the local density measurement.

In order to measure these properties, we modify the pair finder algorithm presented in \citet{Puskas_2025} to find all neighbours around each primary galaxy out to projected physical separations of 1 Mpc, satisfying the criteria for the mass ratio ($\mu \geq 0.25$) and the pair probability ($\int \mathrm{PPF}(z) dz > 0.1$). 

\subsection{Control sample}
\label{sec:control sample}

\begin{figure*}
 \includegraphics[width=\linewidth]{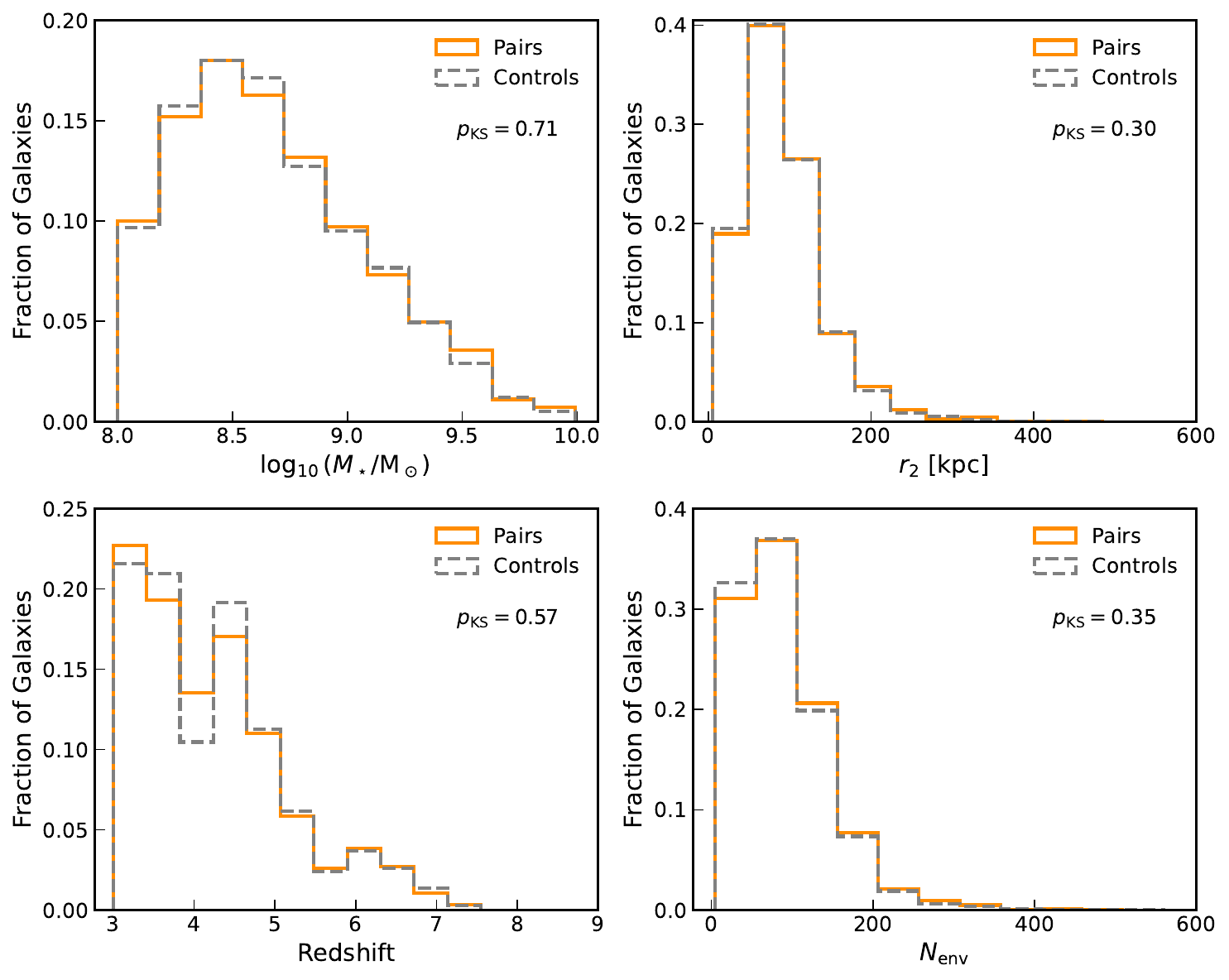}
 \caption{Histograms of the distributions of stellar mass ($M_{\star}$), isolation ($r_2^{\rm pair}$ for paired galaxies and $r_{\rm p}^{\rm ctrl}$ for controls), redshift and local density ($N_{\rm env}$) of the selected galaxy pairs and the matched control sample. The final sample contains 1968 paired galaxies, each matched to five unique controls satisfying the four matching criteria. The corresponding two-sample KS test probabilities are shown in each panel and show no evidence for significant differences in the marginal distributions of the four matching parameters between the two samples, supporting the quality of the control matching.}
 \label{fig:matched_control_distributions}
\end{figure*}

To robustly assess the influence of close companions, it is essential to compare these systems against a carefully constructed control sample. The control matching must be rigorously done, especially at such high redshifts, since most inferred galaxy properties already have large uncertainties, and there are several potential effects that could drive scatter in these measurements. Comparison to appropriately selected control samples has been robustly done at low redshift ($z < 0.3$) by several studies (e.g., \citealt{Ellison_2008, Ellison_2010, Scudder_2012, Patton_2013, Patton_2016, Garduno_2021, Ellison_2022}; and up to $z \sim 3$ in \citealt{Shah_2022}). However, the few existing studies at $z > 3$ \citep[e.g.,][]{Duan_2026} lack such a careful control matching, which is partly due to limits in depth and area in high redshift surveys, but also due to the lower abundance of galaxies with similar properties (e.g., stellar mass and redshift).

In this work, we threfore construct a robust control matching to show the genuine influence of the closest companions. To this end, we match a statistical control sample to each pair system in four parameters: \textit{redshift}, \textit{stellar mass}, \textit{local density}, and \textit{isolation}. We want to compare our selected pairs to galaxies that have similar properties (redshift and stellar mass) and environment, the only difference lying in the presence of a close companion in the former case. The control sample plays an important role, because many physical properties depend on stellar mass and redshift, such as SFR \citep[star forming main sequence, see, e.g.,][]{Speagle_2014, Popesso_2023, Simmonds_2025}, or metallicity (mass-metallicity relation, e.g. \citealt{Nakajima_2023, Curti_2024}). Galaxy properties are also correlated with environment, such as the case of overdensities \citep[e.g.,][]{Helton_2024b} that can also potentially drive Ly$\alpha$ emission \citep[e.g.,][]{Larson_2022, Leonova_2022, Whitler_2024}, as well as quenching \citep[e.g.,][]{Alberts_2022,Baker_2026}. By carefully matching the control sample, we can ensure that any observed differences between galaxy pairs and their controls are not the result of underlying differences in fundamental properties.

The control matching is implemented such that we simultaneously search in a four-dimensional parameter space for the closest match. The initial control pool is identical to the initial sample of primary galaxies described in Section~\ref{sec:galaxy pair selection}, but we exclude all galaxies that belong to the fiducial pair sample, including both primary and secondary members of systems satisfying $r_{\rm p}\leq100$ kpc and $\int_{3}^{9}\mathrm{PPF}(z)dz\geq0.7$. This reduces the initial control-candidate pool from 12903 to 9596 galaxies, ensuring that none of the control candidates is itself a member of a fiducial high-probability pair. For matching in local density, we use our estimates of $N_{\rm env}$ for both the galaxy pair and the control candidate, where in the case of the pair, the closest companion is also included in the calculation of $N_{\rm env}$. Finally, to perform the matching in isolation, we require that the projected distance to the second closest companion ($r_2$) of the galaxy in question (the primary galaxy of the pair system) is similar to the projected distance to the closest companion ($r_{\rm p}$) of the control galaxy. For the control galaxies, $r_{\rm p}$ corresponds to the closest neighbour satisfying the environmental probability threshold, $\int_{3}^{9}\mathrm{PPF}(z)dz>0.1$, and any closer potential major merger companion therefore has an integrated pair probability below this threshold. That is, the magnitude of isolation of the galaxy pair should closely match that of each corresponding control galaxy. See Fig.~\ref{fig:pairs_and_controls} for an example of a galaxy pair and its environment, and three matched controls that agree well in the four properties.

To ensure good matches, further to searching for the closest control in the four-parameter space, we introduce tolerances (see Table~\ref{tab:control} for a summary). We require that the difference between the redshift of the paired galaxy and the control must be smaller than $z_{\rm tol} = 0.3$, where we use the peak of the photometric redshift posteriors or, if available, the spectroscopic redshift. This is significantly higher than, for example, 0.01 used in \citet{Patton_2016}, where they only use spectroscopic redshifts. In the case of stellar mass, we similarly choose $\log_{10}(M_{\rm tol}/M_{\odot}) = 0.3 \ {\rm dex}$, which represents the maximum allowed difference between a pair and a potential control. For comparison, the median \texttt{Prospector} stellar-mass uncertainty is approximately 0.17 dex for both the paired galaxies and matched controls. The adopted tolerance is therefore moderately larger than the typical measurement uncertainty, while the resulting matches are generally considerably closer, as quantified below. We further require an agreement within 25 per cent between $r_{2}$ of the paired galaxy and $r_{\rm p}$ of the control, that is $|1 - r_{\rm p}^{\rm ctrl}/{r_{2}^{\rm pair}}| \leq 0.25$. This is a stronger tolerance than in the case of redshift and stellar mass, which is due to the fact that projected separations can be more reliably measured directly from the photometric images than the former two quantities. In the case of local density, we require agreement within 35 per cent, that is $|1 - N_{\rm env}^{\rm ctrl}/N_{\rm env}^{\rm pair}| \leq 0.35$, due to the more uncertain nature of the environment selection with the minimum pair probability requirement of $\int \mathrm{PPF}(z) dz > 0.1$.

\begin{table}
\centering
\caption{Summary of the four parameters that we use to match a robust control sample to the galaxy pair sample and their chosen tolerances.}
\renewcommand{\arraystretch}{2} 
\resizebox{\columnwidth}{!}{%
\begin{tabular}{lcc}
\hline
Property        & Symbol & Matching tolerance \\
\hline
Redshift      & $z$                               & $\Delta z \leq 0.3$\\
Stellar mass  & $\log_{10}(M_\star/{\rm M_\odot})$& $\Delta \log_{10}(M_\star/{\rm M_\odot}) \leq 0.3$\\
Isolation     & $r_2$~[kpc]                       & $\left| 1 - \frac{r_{\rm p}^{\rm ctrl}}{r_{2}^{\rm pair}}\right| \leq 0.25$\\
Local density & $N_{\rm env}$                     & $\left| 1 - \frac{N_{\rm env}^{\rm ctrl}}{N_{\rm env}^{\rm pair}} \right| \leq 0.35$\\
\hline
\end{tabular}%
}
\label{tab:control}
\end{table}

We perform the matching to find the 5 closest unique controls in the four-dimensional parameter space within the respective tolerances for each paired galaxy. If fewer than five unique controls satisfy all four matching criteria, we exclude that paired galaxy from the subsequent analysis. A given control galaxy may be matched to more than one different pair system, but the same control is never repeated within the set of five controls assigned to an individual paired galaxy. To assess the quality of the resulting control matching, we use two-sample Kolmogorov--Smirnov (KS) tests for each of the four matching parameters, adopting $p_{\rm KS}\geq0.3$ as a benchmark for satisfactory agreement between the pair and control distributions.

Of the 2095 paired galaxies entering the control matching, 1968 have five unique non-pair controls satisfying all four criteria, while 127 galaxies (6.1 per cent) are excluded because fewer than five suitable controls can be found. The resulting matched catalogue therefore contains 9840 pair--control assignments. The KS test probabilities between the pair and matched-control distributions are 0.71, 0.57, 0.30, and 0.35 for stellar mass, redshift, isolation, and local density, respectively (Fig.~\ref{fig:matched_control_distributions}), showing no evidence for significant differences in these marginal distributions. Since a control galaxy can be used for more than one different pair, we additionally compare every paired galaxy directly to the median properties of its five controls. This gives KS probabilities of 0.60, 0.40, 0.60, and 0.63 for stellar mass, redshift, isolation, and local density, respectively. The median pair--control differences are only $\Delta\log_{10}(M_\star/{\rm M_\odot})=0.007$ dex and $\Delta z=0.000$, while the median ratios are $r_{2}^{\rm pair}/r_{\rm p,med}^{\rm ctrl}=1.013$ and $N_{\rm env}^{\rm pair}/N_{\rm env,med}^{\rm ctrl}=1.019$. Thus, the actual pair--control agreement is substantially tighter than the maximum tolerances allowed during the matching.

We note that requiring five unique non-pair controls introduces a modest additional selection on the pair sample, with 127 out of 2095 systems being excluded. These rejected systems preferentially occur at smaller projected separations, higher redshifts and stellar masses, and lower local densities than the retained sample. We therefore note that the final matched sample is somewhat less complete in these regimes, particularly at the smallest separations and highest redshifts.

\section{Enhanced Star Formation in Galaxy Pairs}
\label{sec:Enhanced Star Formation in Galaxy Pairs}

\begin{figure*}
 \centering
 \subfloat{\includegraphics[width=0.5\linewidth]{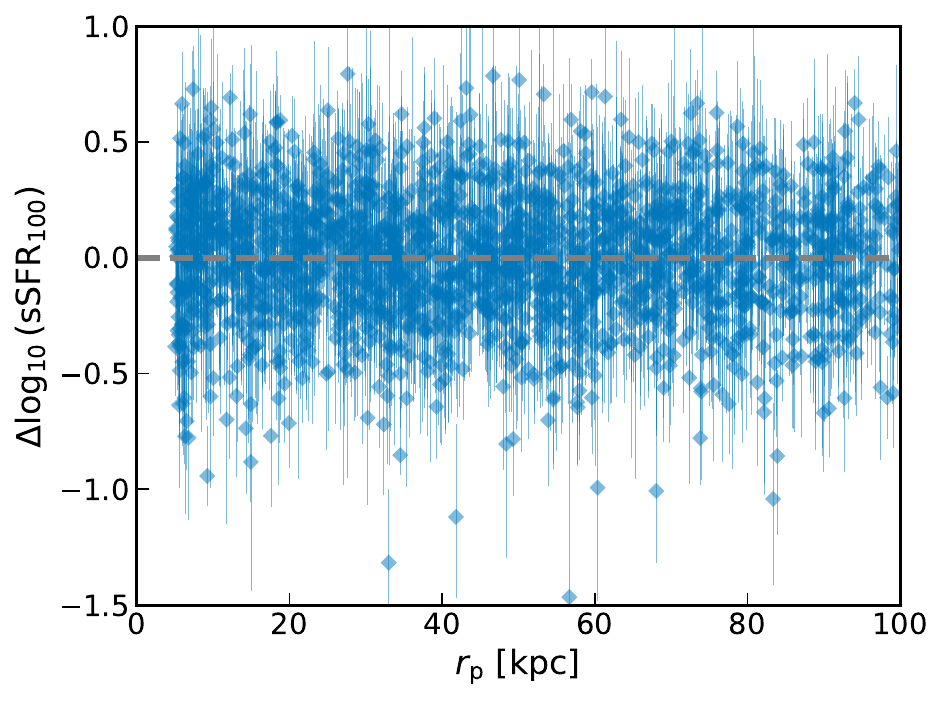}}
 \subfloat{\includegraphics[width=0.5\linewidth]{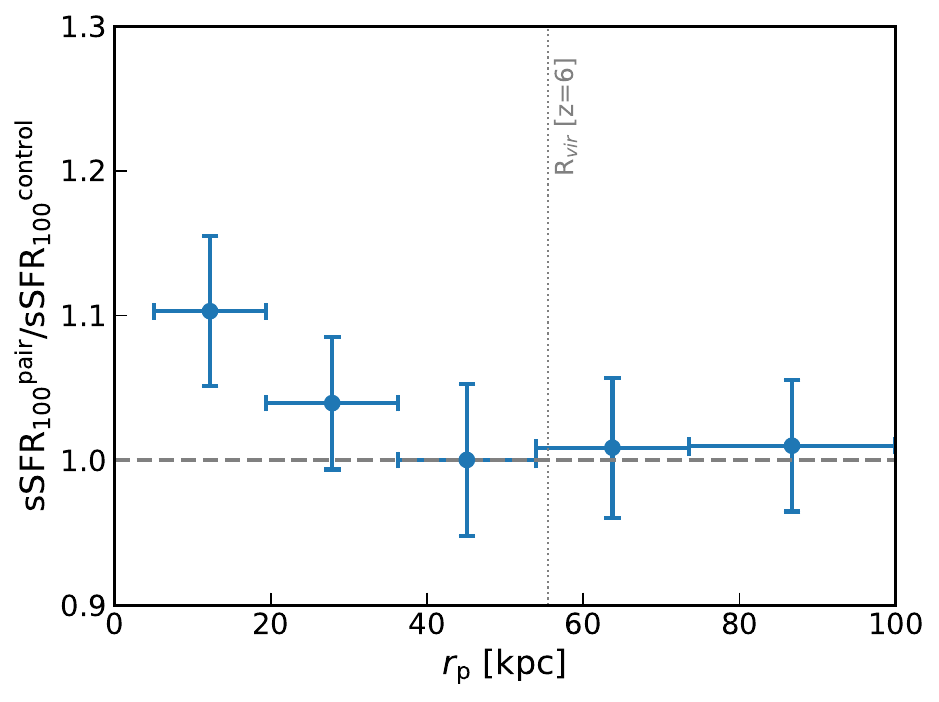}}
 \caption{\textit{Left:} Scatter plot of the sSFR excess of the paired galaxies compared to the median of the controls corresponding to each respective pair as a function of projected pair separation. This is calculated for our final matched sample of high-probability pairs at redshift $z=3-9$ and stellar masses of $\log_{10}(M_\star / {\rm M_\odot}) = [8, 10]$. The uncertainties represent the propagated $16^{\rm th}$ and $84^{\rm th}$ percentiles of the ${\rm SFR_{100}}$ posteriors from \texttt{Prospector}. The dashed grey line represents no excess compared to the control median. \textit{Right:} Bin plot of the ratio between the sSFRs of the paired galaxies and the control medians, where the adaptive bins contain approximately the same number of galaxies. The plotted points represent the median values of the individual ratios (${\rm sSFR_{100}^{pair}}/{\rm sSFR_{100,med}^{ctrl}}$) falling within each respective bin, and the associated uncertainties are the standard errors on the median of the points in the bin. At the smallest projected separations of $r_{\rm p} \lesssim 20 \ {\rm kpc}$, we measure a modest positive sSFR excess of a factor of $1.10 \pm 0.05$, which is consistent with a moderate enhancement of star formation associated with close interactions. The virial radius of dark matter halo at $z=6$ for a typical galaxy in our sample is also displayed.}
 \label{fig:sSFR_enhancement}
\end{figure*}

In this section, we investigate whether the presence of a close companion influences the physical properties of galaxies at $z=3-9$. We provide a comprehensive view of how mergers and interactions affect star formation, both instantaneously and over longer evolutionary timescales. We begin by quantifying the enhancement of star formation in paired systems relative to their matched controls, focusing on specific star formation rate (sSFR) as a function of projected separation (Section~\ref{sec:sSFR distances}). We then explore how these results depend on the timescale over which SFRs are measured (Section~\ref{sec:SFR timescales}), followed by an examination of possible trends with redshift and stellar mass (Section~\ref{sec:Redshift and stellar mass dependence}), as well as studying the environmental dependence (Section~\ref{sec:Environmental dependence}). To gain further insight into the dynamical stage of the interactions, we analyse the reconstructed SFHs of close pairs (Section~\ref{sec:SFH analysis}). 

\subsection{Star formation as a function of pair separation}
\label{sec:sSFR distances}

First, we focus on whether the presence and projected separation of the closest companion have any effect on the star formation of the primary galaxy. Since the SFR has been found to be directly dependent on the stellar mass of the galaxy, following the star-forming main sequence \citep{Speagle_2014, Popesso_2023, Clarke_2024, Cole_2025, Simmonds_2025}, and our galaxy pair sample contains a large range of stellar masses ($\log_{10}(M_\star/{\rm M_\odot})=8-10$), we rather focus on the sSFR. We directly compare the sSFR of the paired galaxies and the median of their corresponding statistical controls, as a function of separation. 

Throughout this work, unless stated otherwise, the control value of a given parameter for each paired galaxy refers to the median value of its five matched control galaxies. This comparison measures the relative offset between paired galaxies and otherwise similar controls, which are closely matched in stellar mass, redshift, and environment. We present in Section~\ref{sec:SED modelling} how we obtain the sSFRs at different timescales, and we choose as our fiducial measure the ${\rm sSFR_{100}}$ (sSFR averaged for the past 100 Myr lookback time). We discuss other choices of timescales for the sSFR measurements and their effect on the results in Section~\ref{sec:SFR timescales}. For the individual measurements shown in the left panel of Fig.~\ref{fig:sSFR_enhancement}, we define the offset sSFR from the control as
\begin{equation}
    \Delta\log_{10}({\rm sSFR_{100}}) = \log_{10}({\rm sSFR_{100}^{\rm pair}}) - \langle \log_{10}({\rm sSFR_{100}^{\rm ctrl}})\rangle_{\rm med},
\end{equation}
where $\langle \cdots \rangle_{\rm med}$ represents the median over the five unique non-pair control galaxies. For the binned measurements in the right panel and throughout the subsequent analysis, we instead present the direct ratio between the value for each paired galaxy and the median value of its controls, ${\rm sSFR_{100}^{pair}}/{\rm sSFR_{100,med}^{ctrl}}$, as this provides a more intuitive measure of the fractional enhancement or suppression.

The left panel of Fig.~\ref{fig:sSFR_enhancement} shows the offset of the individual paired galaxies compared to the control median, which is represented by the grey dashed line, as a function of projected separation. While this figure shows the overall scatter and the uncertainties of the individual data points, the panel on the right-hand side of Fig.~\ref{fig:sSFR_enhancement} is visually easier to interpret, as it shows the ratio of the two quantities with the data binned. We apply an adaptive binning method where the width of each bin (represented by horizontal error bars) is defined such that they contain approximately an equal number of points. The plotted points represent the median value of the distributions of the individual ${\rm sSFR_{100}^{pair}/sSFR_{100}^{ctrl}}$ ratios falling within each bin. Vertical error bars represent the standard error on the median for the data point within each respective bin. As is already evident from the figure, the sSFR of the paired galaxies is consistent with the control median when the closest companion is at wider projected separations. At the smallest separations, $r_{\rm p} \lesssim 20 \ {\rm kpc}$, we find a modest sSFR excess of $10 \pm 5 \%$, while the measurements at larger separations are consistent with the control median within the uncertainties. Given the relatively small amplitude and uncertainty of this excess, we regard it as suggestive evidence for enhanced star formation associated with the presence of a close companion, rather than a strong detection of a continuous radial trend. We note that the separations at which we detect sSFR enhancement are smaller than the virial radius of the dark matter halo at $r_{\rm vir} \simeq55.4$~kpc (assuming a stellar mass-halo mass relation by \citealt{Behroozi_2019}) of a typical galaxy in our sample (i.e., with $10^9~{\rm M_\odot}$ at $z=6$), providing a useful physical scale for comparison. As discussed in Section~\ref{sec:control sample}, requiring five unique controls removes a modest fraction of the original pair sample, with the excluded systems preferentially occurring at smaller projected separations, and this selection should therefore be kept in mind when interpreting the smallest-separation bin. For additional checks on these results, see Appendix~\ref{sec:additional checks}.

\begin{figure}
 \includegraphics[width=\columnwidth]{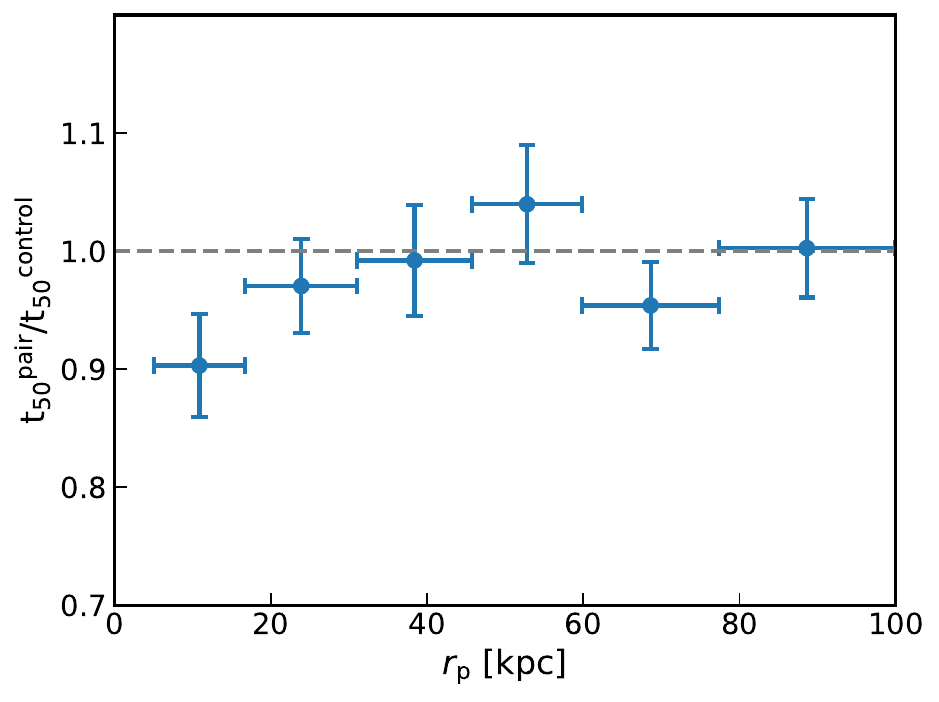}
 \caption{Ratio of the stellar age of galaxies in pairs versus controls as a function of projected pair separation. The smallest-separation bin at $r_{\rm p} \lesssim 20$~kpc shows a lower stellar age for paired galaxies, with $t_{50}^{\rm pair}/t_{50}^{\rm ctrl}=0.90\pm0.04$, while the measurements at larger separations are generally consistent with unity. This suggests more recent stellar mass assembly in the closest pairs, in agreement with the enhanced sSFR found at small projected separations (Fig.~\ref{fig:sSFR_enhancement}).}
 \label{fig:t_50}
\end{figure}

Next, we look at the stellar age (denoted as $t_{50}$) as a function of projected separation. This is the lookback time (relative to the redshift of the galaxy) at which 50$\%$ of the stellar mass of the galaxy has formed and is estimated with \texttt{Prospector} (see Section~\ref{sec:SED modelling}). A smaller $t_{50}$ therefore indicates that a larger fraction of the stellar mass was assembled more recently. As in the case of sSFR enhancement, we calculate the ratio of the stellar ages of the pairs and the median ages of the controls and plot them in adaptive bins as a function of projected separation. As shown in Fig.~\ref{fig:t_50}, the smallest-separation bin has a lower $t_{50}$ compared to the controls, with $t_{50}^{\rm pair}/t_{50}^{\rm ctrl}=0.90\pm0.04$, while the measurements at larger separations are generally consistent with unity. This is consistent with the enhanced sSFR observed at the smallest projected separations (Fig.~\ref{fig:sSFR_enhancement}) and suggests that paired galaxies at these separations assembled a larger fraction of their stellar mass more recently than their matched controls. Since $t_{50}$ is partially degenerate with the recent SFR, we regard this as supporting evidence for the sSFR enhancement rather than a fully independent indication of interaction-induced star formation. Furthermore, beyond this being an effect of the elevated SFR, the primary galaxy could also grow by ex situ accretion of stars (e.g., tidal shredding of a dwarf, as seen in MW tidal streams).

\subsection{SFR timescales}
\label{sec:SFR timescales}

In this section, we examine the sSFRs measured at various averaging timescales and the behaviour of the sSFR enhancement as a function of projected separation, depending on the choice of timescale. \texttt{Prospector} enables us to measure the SFR of a galaxy on different averaging timescales, which, for the purpose of this study, are defined at 5 Myr, 10 Myr, 50 Myr, 100 Myr, and 2000 Myr of lookback time. These timescales provide sensitivity to star formation variability over different temporal baselines and therefore allow us to test whether close interactions measurably elevate recent star formation on specific timescales \citep{Iyer_2020, Tacchella_2020}.

\begin{figure}
 \includegraphics[width=\columnwidth]{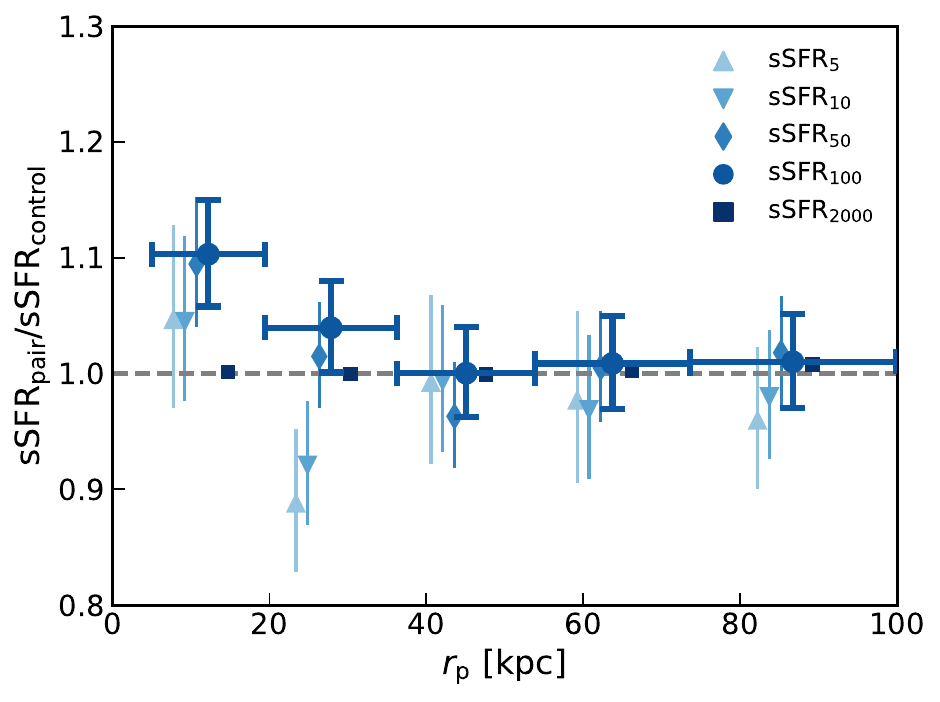}
 \caption{Median sSFR enhancement in paired galaxies relative to controls as a function of projected separation, for averaging timescales of 5, 10, 50, 100, and 2000 Myr. Points are colour--coded from light to dark blue with increasing $t_{\rm avg}$. Adaptive bins contain approximately equal numbers of galaxies, and vertical error bars show the standard error on the median. At the smallest separations, the sSFR excess is clearest for $t_{\rm avg}=100$ Myr (our fiducial choice), with a similar but weaker excess at 50 Myr. At 5--10 Myr the measurements are consistent with the controls, while at 2000 Myr the pair--control ratio remains consistent with unity at all separations.}
 \label{fig:timescales}
\end{figure}

We perform a similar analysis as in Section~\ref{sec:Enhanced Star Formation in Galaxy Pairs} for the sSFRs derived at different timescales and show our results in Fig.~\ref{fig:timescales}. The markers from fainter to darker blue represent sSFR ratios at increasing averaging timescales from 5 to 2000 Myr. As defined in Section~\ref{sec:sSFR distances}, the control value for each paired galaxy is the median of its five matched controls. We only plot the horizontal error bars that represent the width of the adaptive bins (containing approximately equal number of paired galaxies) for ${\rm sSFR_{100}}$, which is our fiducial measure for sSFR. As before, the vertical error bars represent the standard error on the median. In the case of $\rm sSFR_{2000}$, the uncertainties become vanishingly small.

\begin{figure*}
 \centering
 \subfloat{\includegraphics[width=0.5\linewidth]{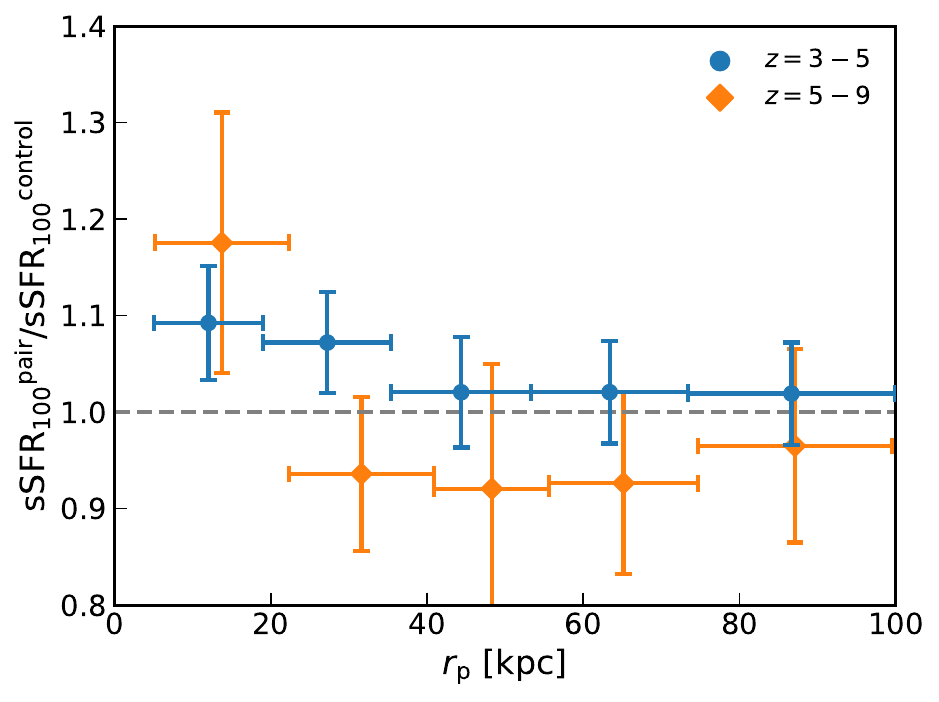}}
 \subfloat{\includegraphics[width=0.5\linewidth]{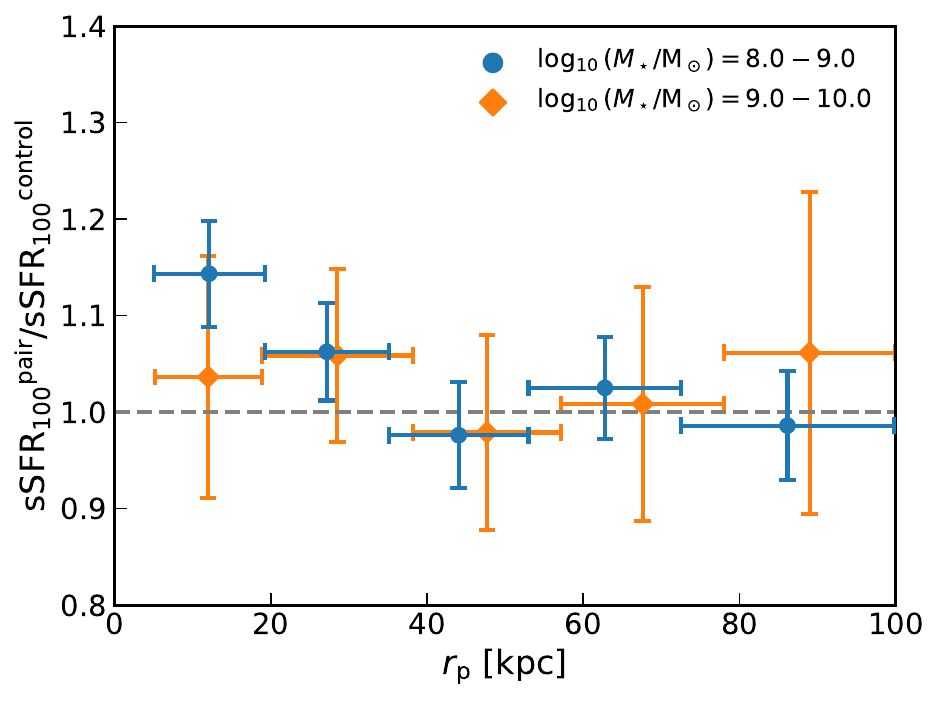}}
 \caption{\textit{Left:} sSFR enhancement measured against projected separation for subsamples divided in redshift, for the ranges $z=[3,5]$ and $z=[5,9]$. At the smallest separations, the higher-redshift subsample shows a somewhat larger sSFR enhancement than the lower-redshift subsample, although the two measurements are consistent within their uncertainties. The contrast between the smallest-separation bin and the following bin is more pronounced at $z=5-9$, while the enhancement varies more gradually with separation at $z=3-5$. \textit{Right:} sSFR enhancement measured against projected separation for subsamples divided in stellar mass, for the ranges $\log_{10}(M_\star / {\rm M_\odot})=[8.0,9.0]$ and $\log_{10}(M_\star / {\rm M_\odot})=[9.0,10.0]$. The lower-mass subsample shows a clearer enhancement at the smallest separations, while the higher-mass measurements have substantially larger uncertainties. The difference therefore provides tentative evidence for a stellar-mass dependence of the enhancement.}
 \label{fig:redshift_mass_dependence}
\end{figure*}

As we decrease $t_{\rm avg}$ from 100 to 50, 10, and 5~Myr, the measured excess in sSFR weakens. By 5--10~Myr, the sSFR `excess' compared to the control median has larger scatter and is statistically consistent with zero, even at the smallest separations ($r_{\rm p}<20$~kpc). This behaviour is consistent with the interaction-related enhancement being most apparent on $\sim$50--100~Myr timescales, while very short-timescale ($\sim$5--10~Myr) sSFR measurements may be more strongly affected by internal, stochastic variations in star formation, such as bursty star formation regulated by gas inflow and feedback. We also note a moderate decrease below the control value in the 5--10 Myr measurements at $r_{\rm p}\sim25$~kpc. This may reflect short-timescale stochasticity in the SFHs, although given the modest significance of the deviation and the difficulty of constraining such rapid variations from the reconstructed SFHs, we do not interpret it as clear evidence for interaction-induced suppression. On the longest averaging timescale (e.g., $t_{\rm avg}=2000$~Myr), the sSFR--separation trend vanishes. This is expected given the young ages of the Universe across our redshift range. For example, at $z\approx3$, where the Universe is $t_{\rm age}\simeq2.16$~Gyr old, averaging over $\sim2$~Gyr covers nearly the entire available star-formation history and therefore approaches ${\rm sSFR}_{2000}\approx1/(2000~\mathrm{Myr})$. At higher redshifts, where the age of the Universe is even shorter, this averaging timescale similarly probes essentially the full available SFH. Consequently, pairs and controls show no detectable contrast on this baseline, thereby just providing a cross-check of our method.

\subsection{Redshift and stellar mass dependence}
\label{sec:Redshift and stellar mass dependence}

In this section, we investigate whether the subsamples divided by either redshift or stellar mass have the same behaviour as the overall sample. We perform this split into two subsamples separately in the case of redshift and stellar mass, as our limited sample size does not allow for splitting into further subcategories.

First, we divide our sample in redshift into two subsamples, that is, at $z = [3, 5]$ and $z = [5, 9]$. The higher-redshift subsample contains fewer galaxies, reflecting the declining number of selected pairs towards higher redshift. We perform a similar analysis as before, and plot the sSFR enhancement with separation on the left panel of Fig.~\ref{fig:redshift_mass_dependence}. At the smallest separations ($r_{\rm p}<20 \ {\rm kpc}$), the enhancement is somewhat higher for the higher-redshift sample (a factor of $\sim1.18$ compared to the control median) than for the lower-redshift sample ($\sim1.09$). However, the uncertainties of the two measurements overlap, and we therefore do not find statistically compelling evidence for evolution in the overall strength of the interaction-induced sSFR enhancement with redshift. Nevertheless, the contrast between the smallest-separation bin and the following bin appears more pronounced at $z=5-9$, while the enhancement varies more gradually with separation at $z=3-5$. This may indicate a difference in the characteristic separation over which the enhancement becomes important, although the current uncertainties do not allow to draw a strong conclusion. This is also broadly consistent with the recent high-redshift major merger pair analysis of \citet{Duan_2026}, who find similar SFR enhancements in their two redshift intervals.

One possible interpretation of this tentative redshift dependence is that galaxies are more gas-rich at earlier times on average \citep[e.g.,][]{Tacconi_2018, Tacconi_2010, Parlanti_2023}, which may affect their response to close interactions, although the dependence of the relative merger-induced SFR enhancement on gas fraction is not necessarily monotonic \citep[e.g.,][]{Scudder_2015,Fensch_2017}. Furthermore, close pairs at higher redshifts may have a higher probability of being observed during first infall, while the lower-redshift sample may contain a larger fraction of systems that have already undergone their closest encounter and are subsequently observed at wider separations \citep[e.g.,][]{Patton_2013}. This remains a possible interpretation rather than a direct inference from our data, since projected separation alone does not uniquely determine the orbital stage of an interacting system. The characteristic interaction scales may also evolve with redshift, as the virial radii of dark matter haloes decrease towards higher redshift at fixed halo mass \citep[e.g.,][]{Behroozi_2019}.

\begin{figure*}
\centering
\subfloat{\includegraphics[width=0.5\linewidth]{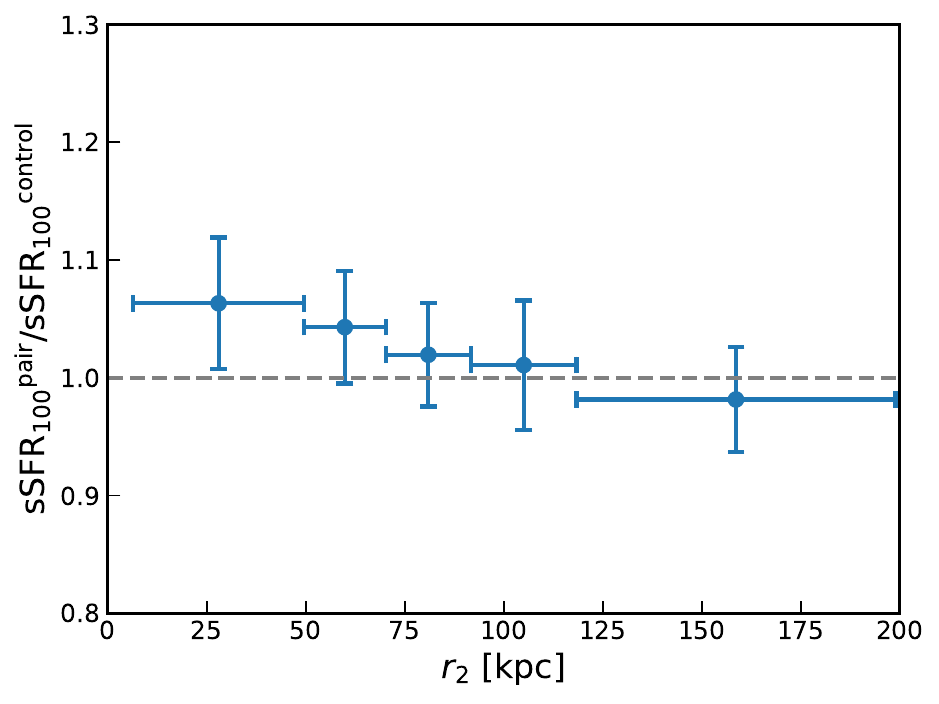}}
\subfloat{\includegraphics[width=0.5\linewidth]{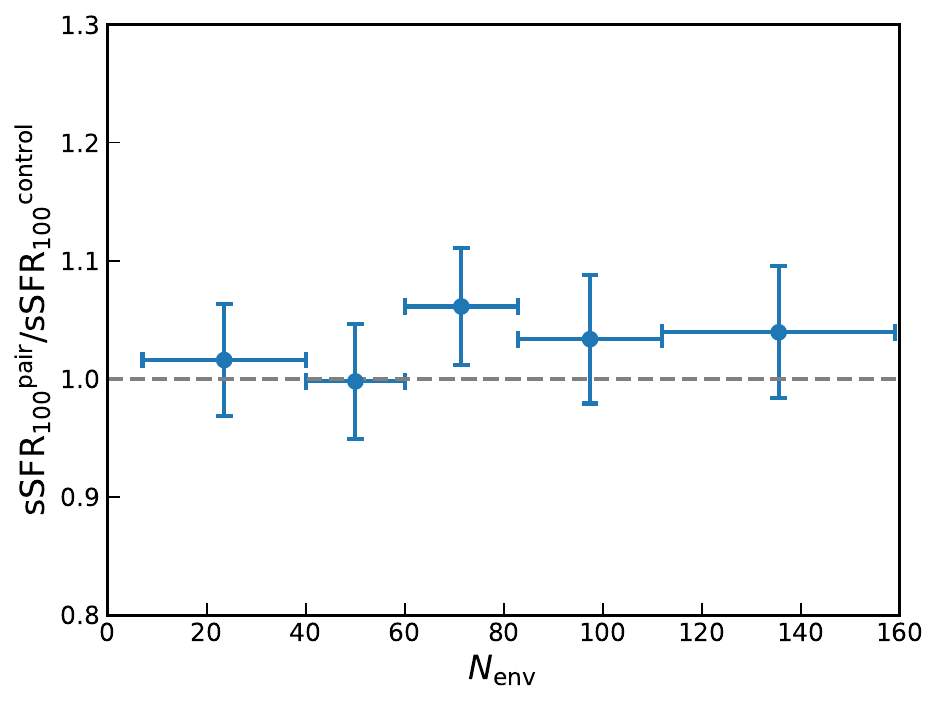}}
\caption{\textit{Left:} sSFR dependence of paired galaxies compared to their controls on the projected separation to the second closest companion ($r_2$). Pairs with smaller $r_2$ show a weak tendency towards higher sSFR ratios, although the measurements are consistent with unity within the uncertainties and we find no statistically significant dependence on isolation. \textit{Right:} sSFR dependence on local density ($N_{\rm env}$), that is given by the number of neighbours within a projected physical separation of 1 Mpc. The pair--control sSFR ratio shows no significant dependence on local density, although there is a weak indication of elevated ratios at $N_{\rm env}\gtrsim60$. All measurements remain consistent with unity within their uncertainties.}
\label{fig:sSFR_environment}
\end{figure*}

Similarly, we divide the sample in stellar mass space into a low- and high-mass regime at $\log_{10}(M_\star / {\rm M_\odot})=[8.0,9.0]$ and $\log_{10}(M_\star / {\rm M_\odot})=[9.0,10.0]$. In this case, the effect of this division is less clear on the sSFR enhancement. Concentrating on the smallest separations ($r_{\rm p}<20$ kpc), the sSFR of lower-mass paired galaxies is more strongly enhanced by a factor of $\sim1.15$ compared to the control median, while the higher-mass measurement is consistent with a substantially weaker enhancement ($\sim1.04$) but has a much larger uncertainty. Given the overlapping uncertainties, we regard this as tentative rather than statistically compelling evidence for a stellar mass dependence. One possible contribution is that lower-mass galaxies generally have higher gas-to-stellar mass ratios \citep[e.g.,][]{Tacconi_2018}, while their shallower potential wells may make them more susceptible to external perturbations. However, given the current uncertainties and the complex dependence of interaction-driven star formation on gas content \citep{Scudder_2015,Fensch_2017}, we cannot identify the physical origin of the tentative mass dependence from these data alone. We also note that the elevated high-mass measurement at the largest projected separations has a large uncertainty and is consistent with the control value, and we therefore do not interpret it as a physical feature.

\subsection{Environmental dependence}
\label{sec:Environmental dependence}

As we already measured the isolation and local density around the pairs as part of the control matching, the environmental dependence of the sSFR naturally comes out of this process. In Fig.~\ref{fig:sSFR_environment} we show how the $\rm sSFR_{100}$ ratio of pairs relative to their controls behaves as a function of the isolation ($r_2$), which we defined as the distance to the second closest companion, and as a function of the local density ($N_{\rm env}$), which is given by the number of neighbours within a projected physical separation of 1 Mpc.

We find a weak indication that the sSFR ratio is higher for pairs with smaller distances to the second closest companion, although all measurements remain consistent with the control value within the uncertainties. This would be in part expected, as it implicitly implies that the distance to the closest companion has to be shorter, i.e., $r_{\rm p} < r_2$. However, $r_2$ is not simply a proxy for $r_{\rm p}$, since the difference between the two distances can vary significantly. Therefore, the weak trend may reflect a combination of the influence of the closest companion and the broader local configuration of neighbouring galaxies. Given the current uncertainties, however, we do not find statistically compelling evidence for an additional dependence of the sSFR enhancement on isolation.

To look further into the influence of the environment on larger scales, we plot the trend of the sSFR ratio against the local density $N_{\rm env}$ in Fig.~\ref{fig:sSFR_environment} (right panel). We find no clear dependence of the pair--control sSFR ratio on local density, with the measurements remaining consistent with unity across the full range of $N_{\rm env}$. There is, however, a very weak indication of systematically elevated sSFR ratios at $N_{\rm env}\gtrsim60$, although the individual measurements are consistent with the control value within their uncertainties and we do not claim a significant trend. Overall, this suggests that the modest sSFR enhancement observed in paired galaxies is not strongly modulated by the larger-scale local density, although a weak enhancement in denser environments cannot be excluded.

\subsection{SFH analysis}
\label{sec:SFH analysis}

In this section, we look at the overall star-formation history (SFH) of the paired galaxies. We use the \texttt{Prospector} outputs for the SFH for each galaxy \citep{Simmonds_2024}. Briefly, we use non-parametric SFH \citep[continuity model; ][]{Leja_2019} and the prior is chosen such that the SFH is divided into eight SFR bins.

We want to investigate the past histories of close pairs, which can be a clue for their present dynamical state. Therefore, we select high-probability \textit{close pairs} with $\int \mathrm{PPF}(z) dz > 0.7$ and $r_{\rm p}<30 \ {\rm kpc}$. We then divide this resulting sample of 627 galaxies into galaxies with SFHs that recently peaked (in the second most recent time bin at $5-10$~Myr), i.e., monotonically rising SFHs (76 galaxies), and with SFHs that already peaked at $> 10$~Myr lookback time (512 galaxies). A further 39 galaxies end up in a category where the SFH first decreases and then increases again, without having a local maximum (peak) in between. We neglect this category of galaxies for the current question.

We note here that when categorising each SFH, we only look at the monotonicity of the SFH beyond the first time bin (at $t_{\rm lookback} > 5$~Myr), because at $t_{\rm lookback} < 5$~Myr the SFR is subject to a larger scatter due to short-term variations \citep{Simmonds_2024, Simmonds_2025}. For the two populations, we then interpolate the SFH of each galaxy onto a common time grid and normalise the SFHs so that the total star formation integrates to 1 (dimensionless; taking the lookback time in units of Gyr). Then we compute the median across all galaxies for the two separate cases and estimate the scatter by the 16--84th percentiles of the distribution of individual SFHs.

\begin{figure}
\includegraphics[width=\columnwidth]{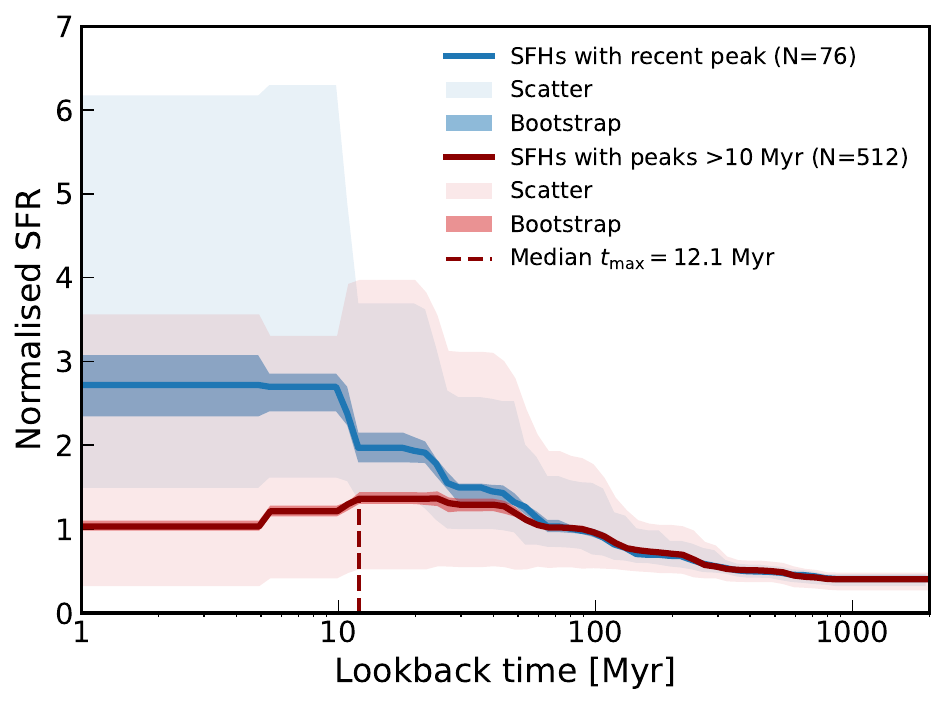}
\caption{Median star-formation histories (SFHs) of galaxies in close pairs with $\int \mathrm{PPF}(z) dz > 0.7$ and separations $r_{\rm p} < 30$~kpc, shown as a function of lookback time. The blue curve and lighter shaded region indicate the median and 16--84$^{\rm th}$ percentile range for galaxies whose SFHs peaked recently, i.e., the SFRs rise monotonically from early times up to the second most recent time bin (median projected separation $r_{\rm p} = 13.19$ kpc). The red curve and lighter shaded region show the same for galaxies with an SFH that peaked at $t_{\rm lookback} > 10$~Myr (median $r_{\rm p} = 16.47$ kpc), i.e., non-monotonically rising SFRs with at least one earlier local maximum. The darker shaded regions show the uncertainties estimated by bootstrapping over the median SFHs. The larger population with earlier SFH peaks (512 galaxies) may include systems that have already undergone a close passage, while the smaller population with recently peaked SFHs (76 galaxies) may include systems experiencing their first approach or pre-coalescence phase.}
\label{fig:sfh_analysis}
\end{figure}

Interestingly, we find that in the case of galaxies that have monotonically rising SFH, their closest companion is at a median projected separation of $r_{\rm p} = 13.19^{+1.44}_{-1.63}$~kpc, while in the case of galaxies that have a secondary peak in their SFH, they have companions at a median separation of $r_{\rm p} = 16.47^{+0.37}_{-0.66}$~kpc. We show the resulting median SFH for the two populations in Fig.~\ref{fig:sfh_analysis} and give the following explanation. One possible interpretation is that the majority of close pairs may have already had their first passage and are observed in their post-pericentre phase. In this scenario, the initial close passage resulted in a starburst and overall rising SFH, which peaked and started to slowly decline as the galaxies moved apart or even passed their apocentre phase (cf. Figure 11 in \citealt{Scudder_2012}). This might have happened multiple times in the history of the pair, but in our overall median SFH calculation, the previous episodes might be `washed out' by the latest rising SFH phase. Only a minority of the selected close pairs have purely rising SFH ($\sim 12$ per cent of the overall cases), which are potentially galaxies that are in their pre-pericentre phase, soon undergoing their first approach, or at their pre-coalescence phase, where, again, previous episodes of close approaches and hence rising and declining SFHs are `washed-out'. This explanation is consistent with the lower median projected separations (13.19 kpc) of galaxies with monotonically rising SFHs compared to the larger separations (16.47 kpc) of the galaxies with secondary peaks in their SFH, which could on average be further away from the primary galaxies either while still receding or approaching again. However, projected separation and the reconstructed SFH alone do not uniquely determine the orbital stage of an individual pair.

\section{Interaction driven AGN activity and Ly\texorpdfstring{$\alpha$}{alpha} emission}
\label{sec:merger driven AGN and LAE}

In this section, we investigate whether close galaxy pair interactions trigger AGN activity and Ly$\alpha$ emission in the studied redshift, stellar mass, and separation limits.

The AGN used in this work have been identified in the following two works: \citet{Juodzbalis_2026} and \citet{Scholtz_2025} for the type-1 and type-2 selection, respectively. The type-1 AGN are identified based on the broad components in the H$\alpha$ and H$\beta$ emission lines without corresponding kinematical components in the [OIII]$\lambda\lambda$5007,4960, hence ruling out an outflow origin. In total, this work yields a clean selection of 35 type-1 AGN in the GOODS-South and -North fields. The type-2 AGN are selected based on a variety of emission line diagnostics in the first two tiers of the JADES survey: goods-s-deephst and goods-s-ultradeep. The emission line diagnostics used were the BPT \citep{Baldwin_1981} and VO87 diagrams \citep{Veilleux_1987}, with selection criteria modified for high-z, the HeII$\lambda$4686 diagrams \citep{Shirazi_2012}, UV emission line diagnostics \citep{Hirschmann_2019} and presence of high ionization lines such as [NeIV]$\lambda$2424 and [NeV]$\lambda3420$. In total, this selected 42 unique type-2 AGN in the regions of the GOODS-South field covered by goods-s-deephst and goods-s-ultradeep. We note that these spectroscopic selections do not provide a complete census of AGN activity, since low-luminosity AGN can be outshone by their host galaxies and several commonly used narrow-line diagnostics become less efficient at identifying AGN at high redshift \citep{Scholtz_2025,Juodzbalis_2026}. Therefore, the number of identified AGN should be regarded as a lower limit to the underlying AGN population.

\begin{figure*}
 \centering
 \subfloat{\includegraphics[width=0.5\linewidth]{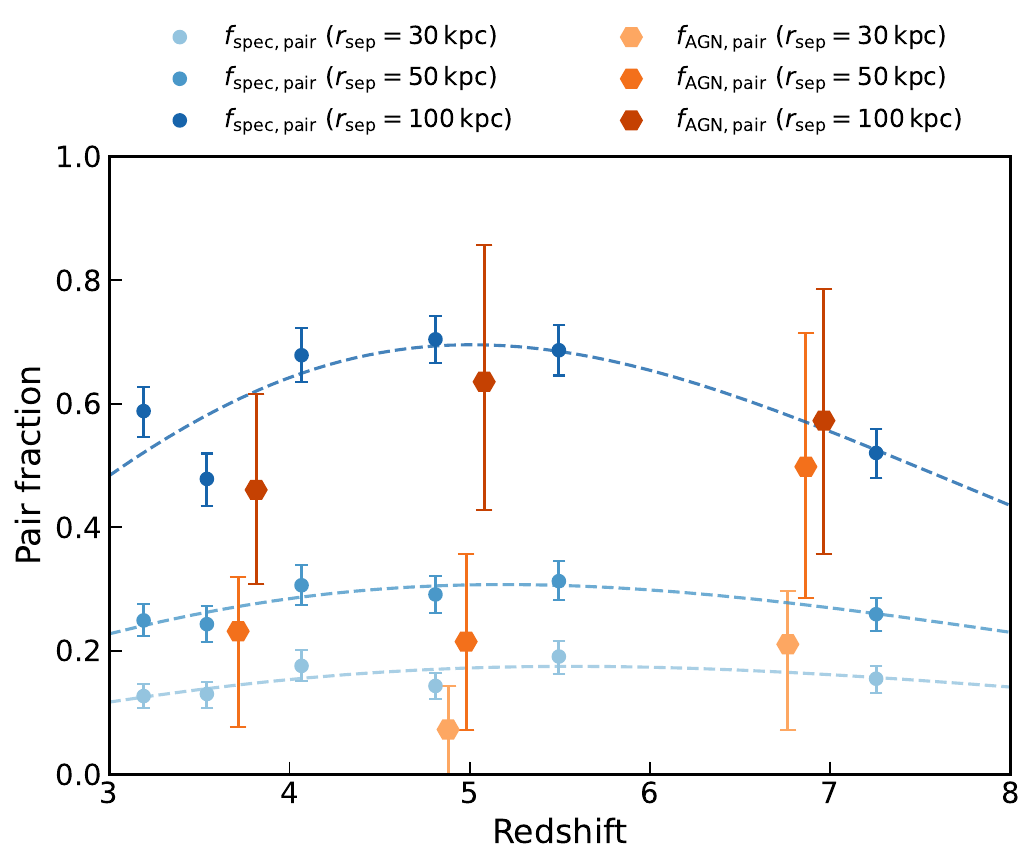}}
 \subfloat{\includegraphics[width=0.5\linewidth]{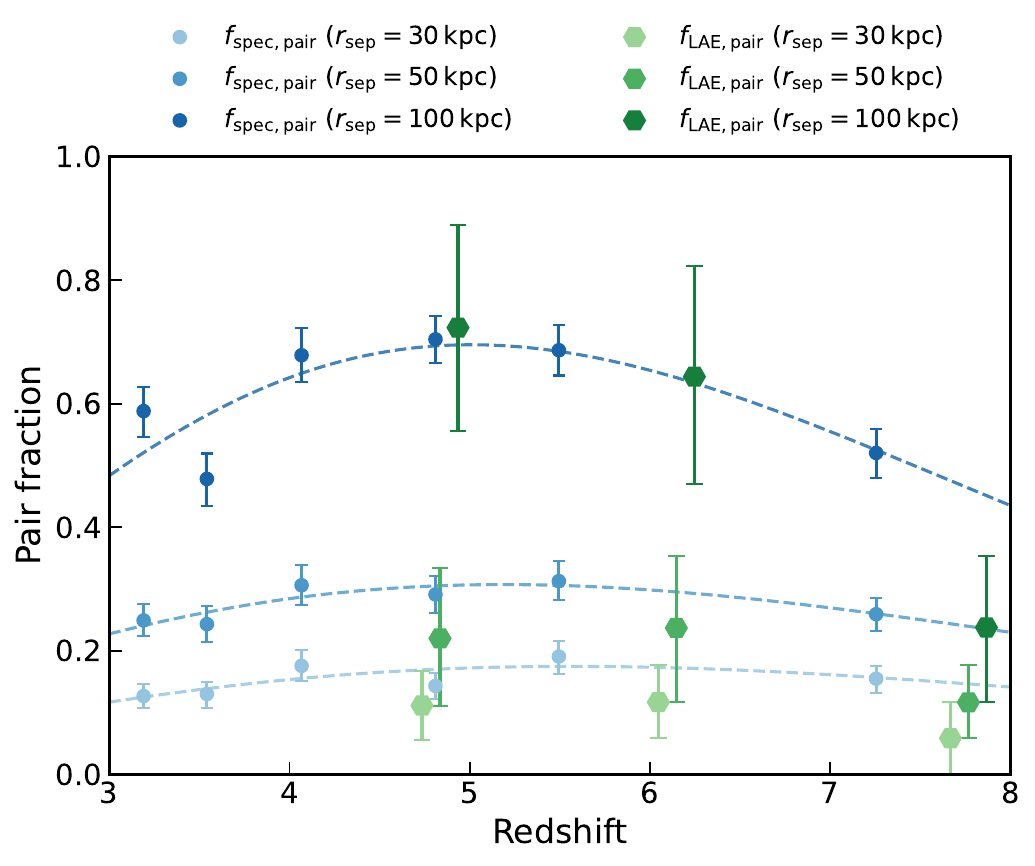}}
 \caption{Pair fraction of AGN (left) and LAE (right) as a function of redshift. Both panels show the pair fraction at different maximum projected separation limits for the subsample with spectroscopic redshifts, indicated by blue circles from fainter to darker shades for increasing separation limits of 30, 50, and 100 kpc. The bin widths are chosen such that each bin contains approximately the same number of objects, and the error bars are estimated by bootstrapping. \textit{Left:} fraction of type-1 and type-2 AGN found in pairs at different maximum projected separations compared to the spectroscopic pair fraction. The paired AGN fraction is generally consistent with the spectroscopic pair fraction within the uncertainties, with a detectable excess (factor of $\approx\!2$) at $r_{\rm p}<50$~kpc in the highest redshift bin ($z>5.5$). At the separations probed here ($r_{\rm p}>5$~kpc), we therefore do not detect a systematic excess of AGN in pairs, although stronger AGN triggering may occur at smaller separations or during coalescence. \textit{Right:} a similar comparison, where the fraction of LAEs in pairs is consistent with the spectroscopic pair fraction at $z<7$ and lower in the highest redshift bin for all separation limits. This may indicate that galaxy interactions are not sufficiently efficient at the separations probed here to enhance Ly$\alpha$ photon production and escape, which may become more relevant at closer separations.}
 \label{fig:AGN_LAE}
\end{figure*} 

The Ly$\alpha$ emitter (LAE) catalogue from \citep{Jones_2025a} consists of all galaxies in JADES DR3 \citep{D'Eugenio_2025}, with secure spectroscopic redshifts (from visual inspection) of $z_{\rm spec}>4$. This redshift cut was chosen so that Ly$\alpha$ would be observable with NIRSpec PRISM/CLEAR, and resulted in a sample of 795 galaxies (150 of which are confirmed LAEs). We refer the reader to Table 1 of \citet{Jones_2025a} for JADES tier separation, Fig. 1 for spatial distribution, and Fig. 2 for redshift distribution. Using the DR3 spectra, models were fit to the PRISM and medium-resolution grating data (if available) that included continuum and strong line emission (including Ly$\alpha$). LAEs in this catalogue are those which show significant Ly$\alpha$ emission (i.e., $S/N>3$) in either PRISM/CLEAR or G140M/F070LP.

Since the identification of both AGN and LAEs needs spectroscopic observations, we have to perform our pair analysis on a spectroscopic sample in order to account for the JADES spectroscopy selection function self-consistently. Specifically, we only focus on pairs that have at least one spectroscopic measurement (either the primary galaxy and/or its companion). For the AGN and LAE pair fractions, the AGN or LAE itself necessarily has a spectroscopic measurement, while its companion does not necessarily require one and can enter the pair selection through its photometric-redshift probability distribution. Thus, we measure the fraction of spectroscopically identified AGN or LAEs associated with galaxy pairs, rather than requiring spectroscopy for both members of each pair. Therefore, the resulting galaxy pair fraction (number of pairs divided by the number of primary galaxies in the sample) will be different by definition compared to the results reported in \citet{Puskas_2025}, where photometric redshifts were mostly used for close pair selection. We recalculate the galaxy pair fraction for this spectroscopic-only catalogue for different maximum pair-separation limits (30, 50, and 100 kpc). In this case, we define the pair fraction as the number of galaxies with spectroscopic redshifts being in pairs (either being the primary or the secondary galaxy within the pair) divided by the total number of galaxies having spectroscopic redshifts within a given redshift bin (both being within the stellar mass range of $\log_{10}(M_\star/{\rm M_\odot})=8-10$). In total, we have 3249 objects with spectroscopic redshifts in GOODS-South and GOODS-North that remain in our catalogue after the initial sample selection (see Section~\ref{sec:galaxy pair selection}), which is further reduced to 1928 for the redshift and stellar mass range of interest. We then divide our redshift range of $z=3-9$ into six bins with adaptive widths so that each bin approximately contains an equal number of galaxies. For the pair selection, we use the same parameter choices as for our fiducial pair selection in Section~\ref{sec:galaxy pair selection}, that is $\int {\rm PPF(z)} dz >0.7$. The resulting pair fractions ($f_{\rm spec, pair}$) are plotted in both panels of Fig.~\ref{fig:AGN_LAE} for the three different separation limits (from fainter to darker shades of blue for increasing separation limits), and their uncertainties are estimated by bootstrapping \citep{Efron_1979, Efron_1981}. We also fit a power law-exponential curve of the form
\begin{equation}
 f_{\rm pair}(z)=f_0(1+z)^m \exp[\tau(1+z)],
\end{equation}
where $f_0$, $m$, and $\tau$ are free parameters \citep{Puskas_2025}, to the pair fractions for the three cases for easier readability. We note here that $f_{\rm spec, pair}$ at $r_{\rm p} < 30$~kpc agrees well with the galaxy pair fraction reported by \citet{Puskas_2025} using both photometric and spectroscopic samples.

We proceed in a similar way to calculate the fraction of AGN in pairs ($f_{\rm AGN, pair}$) and LAE in pairs ($f_{\rm LAE, pair}$). We first match the type-1 and type-2 AGN catalogue to the NIRCam footprints used for this analysis and find that 64 AGN remain in our catalogue after the initial sample selection (Section~\ref{sec:galaxy pair selection}). In the case of LAEs, 132 objects remain in the initial sample catalogue. The fractions are then calculated as the ratio of AGN or LAEs in pairs (either primary or secondary galaxies) to the total number of AGN or LAEs in our sample. If an AGN or LAE appears as both a primary and a secondary galaxy in different pair systems, these occurrences are treated as separate pair associations, since they correspond to distinct pair configurations rather than duplicate entries of the same system. Due to the low number of sources above $z>8$ (2 AGN and 7 LAEs), we focus on the redshift range of $z=3-8$ in this section. We define the bins similarly by the adaptive method, but reduce their number to three due to the lower number count of these objects.

The fraction of AGN in pairs has larger uncertainties (estimated by bootstrapping) compared to the pair fractions of galaxies in pairs with spectroscopic redshifts ($f_{\rm spec, pair}$), which is mainly due to the much lower sample size of AGN (see Fig.~\ref{fig:AGN_LAE}). At the largest separation limit of $r_{\rm p} = 100$~kpc, the values are comparable and agree well within the uncertainty ranges, meaning that there is no detectable excess in the number of AGN compared to other galaxies in pairs. As we go to a lower separation limit of $r_{\rm p} = 50$~kpc, we only find a significant excess (factor of $\sim\! 2$) in $f_{\rm AGN, pair}$ compared to $f_{\rm spec, pair}$ at the highest redshift bin of $z=6.8 \pm 1.2$. Finally, at the closest separations of $r_{\rm p}<30$~kpc, $f_{\rm AGN, pair}$ is generally consistent with $f_{\rm spec, pair}$, except at $z=3.6 \pm 0.6$ where we find no paired AGN satisfying the selection criteria (therefore we omit this point from Fig.~\ref{fig:AGN_LAE}). In terms of the total AGN sample, we find that for the entire redshift and stellar mass range, at $r_{\rm p} < 100$~kpc, $f_{\rm AGN, pair} = 0.44 \pm0.07$ when each AGN is counted once if it is present in at least one pair (either as a primary or secondary), and $f_{\rm AGN, pair} = 0.56 \pm0.10$ when distinct pair associations are counted separately, such that an AGN appearing as a primary in one system and a secondary in another contributes to both associations. This is in good agreement with the AGN fraction of $f_{\rm AGN, pair}=0.52^{+0.17}_{-0.24}$ reported by \citet{Duan_2026}, who select AGN by the spectroscopic BPT diagnostic method and photometric AGN SED templates from eight deep \textit{JWST} fields at similar separations limits, but using the simpler redshift criterion of $\Delta z <0.3$. The lack of a systematic increase in $f_{\rm AGN,pair}$ towards smaller projected separations differs from some lower-redshift results, where the strongest AGN excess is found during the latest merger stages \citep[e.g.,][]{Ellison_2025}. We discuss possible explanations for this difference in Section~\ref{sec:implications for LAE and AGN}, including the minimum separation probed by our analysis and the short duty cycle of AGN activity.

The fractions of LAEs in pairs ($f_{\rm LAE, pair}$) have similar uncertainties to $f_{\rm AGN, pair}$ due to comparable number counts. At redshift $z<7$, $f_{\rm LAE, pair}$ is consistent with $f_{\rm spec, pair}$ at all three separation limits, with no detectable excess, similar to $f_{\rm AGN, pair}$. In the highest redshift bin at $z=7.7 \pm 1.0$, we find that $f_{\rm LAE, pair}$ is lower than $f_{\rm spec, pair}$ at all separations. Similarly to the AGN case, we calculate the global LAE fraction at $r_{\rm p} < 100$~kpc as $f_{\rm LAE, pair} = 0.42 \pm0.08$ when each LAE is counted once if it is present in at least one pair, and $f_{\rm LAE, pair} = 0.53 \pm0.09$ when distinct pair associations are counted separately. We note that these values are consistent with the global AGN fractions, showing no detectable excess. Interestingly, when comparing the AGN sample to the LAE sample, we find that the fraction of AGN that are also LAEs is $0.12 \pm 0.05$, while the fraction of LAEs that are also AGN is $0.42 \pm 0.17$, with 12 sources being identified as both AGN and LAE that are present in our pair catalogue. In the next section, we discuss the potential physical origins and explanations for the fraction of paired AGN and LAEs.

\section{Discussion}
\label{sec:discussion}

In this section, we discuss possible physical explanations and implications of our results and compare them to works from the literature. We also discuss how our results depend on parameter choices and mention caveats to this analysis that might affect the results.

\subsection{Implications}
\label{sec:implications}

One of the key questions at these redshifts is whether mergers are the primary drivers of intense star formation or if they provide only a modest boost to galaxies already forming stars rapidly due to smoother gas inflow and high gas fractions. High-redshift disk galaxies have significantly higher gas fractions, with $f_{\rm gas} \gtrsim 50 \%$ being common at $z\sim2$ \citep[e.g.,][]{Tacconi_2010, Scoville_2017, Tacconi_2018, Tacconi_2020, Parlanti_2023}. This abundant fuel can lead to violent disk instabilities (VDI), which drive strong gas inflows and nuclear starbursts \citep{Bournaud_2012, Dekel_2014, Zolotov_2015, Tacchella_2016}. The lifecycle of these VDI-induced clumps occurs on short timescales of 5--10 Myr, which aligns with the stochastic, bursty variability we observe in our short-timescale sSFR measurements (see Section~\ref{sec:SFR timescales}). In contrast, the longer-timescale enhancement (50--100 Myr) we detect may reflect the large-scale gas inflow and potential compression induced by the merger interaction itself. Our results -- showing a separation-dependent boost on $t_{\rm avg}\!\sim\!50$--$100$~Myr but not on 5--10 Myr -- therefore are consistent with merger-driven fuelling operating on longer dynamical timescales, while very short-timescale sSFR is dominated by these internal, stochastic processes \citep{Iyer_2020, Tacchella_2020, Simmonds_2025, McClymont_2025}.

In the low-gas fraction case, typical of the local universe, interaction-induced torques efficiently channel gas to the centre of galaxies, fuelling strong starbursts \citep{Barnes_1991, Mihos_1996}. However, in the gas-rich environments of the early universe, this picture changes. Simulations show that when the gas fraction is high, a strong baseline inflow already exists due to internal processes, and the interaction between galaxies does not significantly increase this inflow until just before coalescence \citep{Fensch_2017}. Similarly, other simulations have found that merger-induced starbursts are often short-lived and difficult to distinguish from the high levels of stochastic star formation driven by internal processes \citep{Sparre_2016, Hani_2020}. Our results align well with this gas-rich merger paradigm. The star formation histories of our high-probability galaxy pair sample are consistent with many pairs having already undergone a first pericentric passage, while the strongest enhancement we find occurs at projected separations $r_{\rm p}\!\lesssim\!20$ kpc. The modest amplitude of this enhancement of a factor of $\!\sim\!1.10$ is in good agreement with simulations like those of \citet{Fensch_2017}, which find only a weak SFR boost at the first pericentre passage and a mild elevation at final coalescence. This scenario is further supported by recent direct observations of high-redshift pairs \citep[e.g., ][]{Duan_2026}, which confirm that merger-driven enhancements are typically a factor of $\!\sim\!2$ or less, far weaker than in local universe mergers \citep{Cibinel_2019}.

It has been found by previous works that galaxies above the SFMS \citep[e.g.,][]{Speagle_2014, Faisst_2016, Popesso_2023, Simmonds_2025} at high redshift are almost always associated with major mergers \citep[e.g.,][]{Rodighiero_2011, Cibinel_2019}. The median sSFR enhancement of $f\!\approx\!1.10$ detected by our study essentially represents a modest offset from the SFMS. This shortens the e-folding and doubling times by $\sim9\%$ (i.e., faster recent mass build-up). This modest `simmering' mode of enhancement, seen on $\sim50$--$100$ Myr timescales, is a key mechanism for pushing galaxies up along the mass-SFR plane faster than their isolated counterparts. Mergers, therefore, act as a dual growth engine: they add mass directly via the accretion of the secondary galaxy (see, e.g., \citealt{Puskas_2025}) and indirectly by triggering in-situ star formation that builds the stellar mass of the primary galaxy more rapidly. 

As presented in Section~\ref{sec:Redshift and stellar mass dependence}, we find a somewhat larger enhancement at higher redshifts ($f \approx1.18$, at $z=5-9$) and at lower masses ($f \approx 1.15$, at $\log_{10}(M_\star / {\rm M_\odot})=8.0-9.0$), although in both cases the measurements are consistent with the respective comparison subsamples within their uncertainties. These trends should therefore be regarded as tentative. The possible stronger enhancement at higher redshift could be tied to the evolution of the cosmic gas fraction. As discussed before, galaxies at $z>3$ are overwhelmingly gas-dominated. A merger between two gas-rich systems provides a massive, concentrated fuel reservoir for a starburst \citep{Tacconi_2018}. At lower redshifts, galaxies are more gas-poor, therefore, a similar interaction may result in a weaker SFR enhancement. The dependence of the star formation enhancement on the stellar mass range studied can be explained by the following. More massive galaxies have deeper gravitational potential wells and higher stellar densities, which makes their gas disks more stable against external perturbations. Therefore, it is more difficult for a merger to disrupt a massive disk, whereas a lower-mass and less stable disc is more easily disturbed, leading to a more vigorous starburst \citep{Pearson_2019}. Moreover, low-mass galaxies have shallower potential wells, making it easier for stellar feedback to expel gas. However, the initial burst triggered by a merger can be more intense because there is less pre-existing pressure support in the disc to resist gas inflow and compression \citep[e.g.,][]{Muratov_2015}. Given the overlapping uncertainties between the mass subsamples, however, these physical explanations remain tentative.

\subsection{Implications for Ly\texorpdfstring{$\alpha$}{alpha} emitters and AGN}
\label{sec:implications for LAE and AGN}

Here, we discuss the possible explanations and implications of finding no significant excess in the Ly$\alpha$ emitter fraction among close pairs compared to the overall spectroscopically confirmed galaxy pair population. During the EoR ($z>5.2$), the intergalactic medium (IGM) is substantially neutral, and neutral hydrogen is extremely effective at scattering Ly$\alpha$ photons, making it difficult for the Ly$\alpha$ emission from most galaxies to reach us. Ly$\alpha$ photons can only escape from regions where the surrounding IGM has already been ionized, creating channels or large bubbles of ionized gas. If galaxy pairs were significantly more efficient at producing such ionized regions through their combined ionizing output, we would expect an enhanced LAE fraction among pairs. The absence of such an excess in our results instead suggests that either (i) galaxy pairs do not play a dominant role in creating ionized bubbles large enough to boost Ly$\alpha$ visibility, or (ii) the local environment, including neighbouring galaxies and large-scale overdensities, is the primary driver of Ly$\alpha$ transmission during reionization. 

We note that our analysis probes relatively wide projected separations ($r_{\rm p}=5-100$~kpc). At these scales, although we detect a modest enhancement in star formation at the smallest separations, many pairs may not yet be undergoing strong interactions, so their combined ionizing output is likely weaker than for very close pairs ($r_{\rm p}<5$ kpc). This is further supported by the lack of star formation enhancement on $\sim\!10$~Myr timescales (see Section~\ref{sec:SFR timescales}), which is consistent with the absence of a strong very recent increase in the production of Ly$\alpha$ photons. At the smallest separations (which we do not probe), stronger star formation enhancements and disturbed morphologies could lead to less uniform neutral gas coverage and, consequently, more efficient Ly$\alpha$ escape. Thus, our results do not rule out the possibility that very close pairs ($\lesssim5$~kpc) or compact groups provide particularly favourable conditions for Ly$\alpha$ emission during the EoR, consistent with a patchy or heterogeneous picture of reionization that begins in overdense regions such as those hosting close galaxy pairs and protoclusters \citep{Castellano_2016, Endlsey_2021, Saxena_2023, Whitler_2024, Witten_2024, Witstok_2024}.

In the case of AGN, we find a lack of significant enhancement in $f_{\rm AGN, pair}$ compared to the spectroscopically selected pair fractions. While mergers are theoretically a very effective way to drive gas to the galactic nucleus to fuel a supermassive black hole, the AGN phase itself is thought to be very brief and episodic. This is referred to as the AGN duty cycle model \citep{Schawinski_2015}. An AGN might turn on for a short period ($\!\sim\!0.1-1$~Myr) and then shut off for a much longer period ($\!\sim\!10-100$ Myr) as the fuel is consumed or expelled by feedback \citep{Hickox_2014}. This model aligns with our findings, as we observe a sustained SFR enhancement on a long timescale of $\!\sim\!50-100$~Myr, which reflects the overall duration of the galaxy interaction. However, we see no sSFR enhancement on the very short $\!\sim\!5-10$~Myr timescale. The lack of an AGN trend is consistent with the idea that the AGN accretion phase is a short-timescale phenomenon, similar to the stochastic starbursts. Even if the merger is the ultimate cause of the fuelling, the probability of observing any given pair during the brief moment its AGN is on is low. Furthermore, as discussed in Section~\ref{sec:merger driven AGN and LAE}, our spectroscopic AGN selection does not provide a complete census of AGN activity, particularly at low luminosities, which further limits our sensitivity to a modest interaction-driven excess.

On the other hand, the probability of detecting an AGN is expected to increase during the strongest interaction phases. For example, \citet{Perna_2025} find that 20\%-30\% of their sample of AGN is dual AGN at $z \sim3$ (which is in excess by a factor of $\sim\!3$ relative to expectations by simulations). They explore dual AGN on scales of $3-20$~kpc, suggesting a significant role of galaxy interactions on AGN activity. However, in our selection criterion, we limit our study of galaxy pairs to projected separations $r_{\rm p}>5$ kpc, likely introducing a significant bias against detecting merger-triggered AGN, which happen primarily on scales smaller than what is probed in this study. 

Theoretical models and simulations consistently show that the most efficient period of AGN fuelling occurs during the final stages of a merger \citep[e.g.,][]{Hopkins_2008, Ellison_2011, Koss_2012}. Similarly, \citet{Ellison_2025} find observationally that the AGN excess is strongest immediately following galaxy coalescence. This regime is not probed by our pair selection, providing a natural explanation for why we do not recover a similarly strong enhancement. It is at these small separations ($<5$ kpc) that violent gravitational torques can strip gas of its angular momentum, driving rapid inflows toward the central supermassive black holes. By excluding this final coalescence phase, our study is systematically insensitive to the brief, intense period where AGN are most likely to be triggered, which could explain the lack of a strong correlation between pairing and AGN activity in our sample.

Recently, \citet{Ubler_2024} reported the discovery of a spatially offset (by $\sim\!1$~kpc) AGN at $z = 7.2$, that is likely undergoing a merger with another galaxy (likely hosting another accreting black hole). \citet{Ubler_2026} present a galaxy at a $z\sim5$ hosting three massive black holes (a central one separated by the second at only $\sim\!200$~pc and the third being at $\sim\!2$~kpc), providing direct evidence that SMBH mergers and multi-SMBH systems were already active during the early stages of galaxy assembly. At later times, \citep{Bonaventura_2025} show that at redshift $z \sim 1-3$, AGN are commonly dust-obscured and live in disturbed host galaxies, revealing a strong link between dynamical processes such as mergers and obscured black hole growth.

\subsection{Caveats and parameter choices}
\label{sec:caveats and parameter choices}

\begin{figure*}
\includegraphics[width=\linewidth]{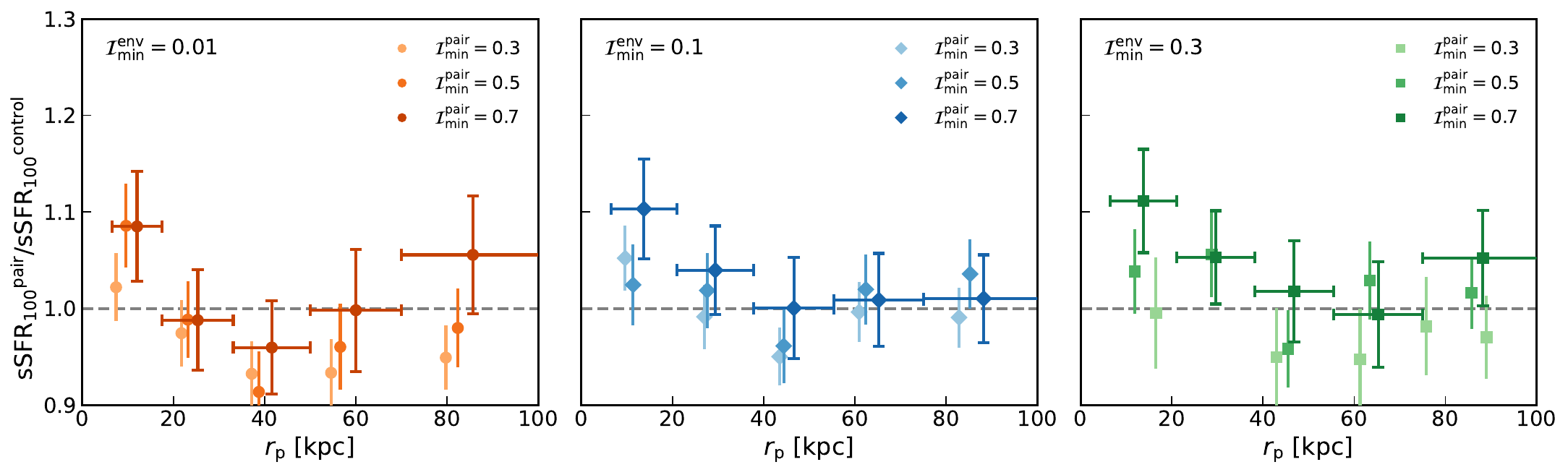}
 \caption{Enhancement in sSFR for paired galaxies versus projected separation for varying pair probability selections. The three panels adopt environmental probability thresholds of $\mathcal{I}_{\rm min}^{\rm env}=0.01$, 0.1, and 0.3, respectively, from more generous to more restrictive neighbour selections. In each panel, the pair-selection threshold $\mathcal{I}_{\rm min}^{\rm pair}$ is varied between 0.3, 0.5, and 0.7, corresponding to increasing pair probability. Our fiducial selection is $\mathcal{I}_{\rm min}^{\rm env}=0.1$ and $\mathcal{I}_{\rm min}^{\rm pair}=0.7$. Across the different parameter choices, we generally recover a similar tendency for elevated sSFR ratios at the smallest projected separations, although the detailed radial behaviour varies and the individual trends are not statistically significant in all cases. Thus, the qualitative indication of a modest enhancement at small $r_{\rm p}$ does not depend strongly on the exact probability thresholds adopted.}
 \label{fig:param_var}
\end{figure*}

In this section, we discuss potential limitations of this study by listing caveats and presenting the effect of different parameter choices on our results.

As described in Section~\ref{sec:galaxy pair selection}, we select major merger galaxy pairs that have mass ratios of $\mu \geq 1/4$. By lowering this value, we could select minor ($\mu \geq 0.1$) or mini ($\mu \geq 0.01$) mergers, however, this would result in a very incomplete sample in stellar mass. Therefore, we focus on only selecting major mergers in this work. For future studies, one could examine the effect on these properties not only by using major merger galaxy pairs but also by considering minor or mini mergers. However, this would require much deeper data for a mass-complete analysis, a challenging task at higher redshifts.

As discussed previously, we study the variation of the sSFR and other properties of (only) the central galaxy in high-probability pairs, as a function of pair separation. We also discuss how this depends on different choices of averaging timescales for the sSFR estimates. One could also look at similar effects on not only the primary or central galaxy of the pair system, but also on the lower mass companion. To perform a similarly robust analysis, control matching has to be done. However, in this case, the control matching would be much more complicated, as the `primary' galaxy of the control system has to be similar (with a similar environment but no massive host?) to the companion galaxy in question, whose environment now is more complicated, being `off-centre' compared to the primary galaxy. The companion, having a lower mass and requiring similar mass controls, would also be more heavily affected by stellar mass incompleteness. Therefore, we only focus on studying the properties of the central (primary) galaxy in this work.

In this analysis, a parameter choice we inevitably made in the case of our probabilistic pair selection is a threshold chosen for the integral of the pair probability function. We defined high-probability pairs as having $\int {\rm PPF}(z) dz > 0.7$ for our fiducial analysis. To detect the second closest companion and measure the local density, we chose a minimum threshold of $\int {\rm PPF}(z) dz > 0.1$. The resulting pairs and their neighbours naturally depend on these parameter choices. Therefore, in this section, we investigate the effect of choosing different threshold values on the resulting sSFR excess measurements.

We perform the same analysis as in Section~\ref{sec:Enhanced Star Formation in Galaxy Pairs}, for three different minimum thresholds for finding neighbours in the local environment, for which we choose the short-hand notation of $\mathcal{I}_{\rm min}^{\rm env} = \int {\rm PPF}(z) dz$. We choose 0.01, 0.1, and 0.3 for $\mathcal{I}_{\rm min}^{\rm env}$, which means a very generous, more restricted, and very pure selection thresholds for neighbours, affected mainly by their redshift posterior distributions. Then, for each $\mathcal{I}_{\rm min}^{\rm env}$, we perform the analysis for three threshold values for finding the closest companion, with the short-hand notation of $\mathcal{I}_{\rm min}^{\rm pair} = \int {\rm PPF}(z) dz$. In this case, we choose higher values (0.3, 0.5, and 0.7), as we want to select higher probability pairs, even if the environmental selection is less restrictive. While we choose the thresholds $\mathcal{I}_{\rm min}^{\rm pair} = 0.7$ and $\mathcal{I}_{\rm min}^{\rm env} = 0.1$ as our fiducial parameters, we generally recover a similar tendency for elevated sSFR ratios at the smallest separations across the different parameter choices, although the detailed radial behaviour varies and the individual trends are not statistically significant in all cases, as shown in Fig.~\ref{fig:param_var}. Therefore, the qualitative indication of a modest enhancement at small separations does not appear to depend strongly on the exact probability thresholds adopted, although its detailed strength and radial dependence remain uncertain. We note that with the `purest' selection using $\mathcal{I}_{\rm min}^{\rm pair} = 0.7$ and $\mathcal{I}_{\rm min}^{\rm env} = 0.3$, our results have more scatter, which is mainly due to the significantly reduced number of pairs satisfying this selection criterion.

\section{Summary \& Conclusions}
\label{sec:conclusion}

In this paper, we study and measure the influence of close companions on different galaxy properties. By careful control matching, we can isolate and study the effects of close galaxy interactions on the physical properties of the paired galaxies, such as star formation enhancement, AGN triggering, and Ly$\alpha$ emission.

We select a sample of high-probability galaxy pairs with projected physical separations of $r_{\rm p} < 100$~kpc at $z=3-9$ and stellar masses $\log_{10}(M_\star / {\rm M_\odot}) = [8, 10]$ from the JADES survey, and compare to a carefully selected control sample by simultaneously matching in redshift, stellar mass, isolation, and local density (within 1 Mpc). We exclude galaxies belonging to the fiducial pair sample from the control pool and require five unique matched controls for each paired galaxy. This matching allows us to isolate statistically the effect associated with a close companion and ensure that any observed differences between pairs and their controls are not the result of underlying differences in fundamental galaxy properties.

We summarise the most important findings of this paper in the list below:
\begin{itemize}

\item We compare the sSFRs of the (primary) galaxies in major merger pairs to their matched controls and find a modest enhancement at the smallest projected separations, reaching an excess of $10 \pm 5$ per cent at $r_{\rm p} \lesssim 20$~kpc, while at larger separations the measurements are consistent with the controls.

\item By measuring the sSFR variation compared to controls with projected separation at different averaging timescales, we find that the interaction-related enhancement is most apparent at timescales of $50-100$~Myr. In contrast, at shorter timescales of $5-10$~Myr, sSFRs are mainly affected by internal processes that induce short-time variability, such as feedback or the stochastic nature of star formation.

\item By splitting our sample in redshift space, we find a somewhat stronger excess of $\times 1.18$ in sSFR at $r_{\rm p} \lesssim 20$~kpc at higher redshifts ($z = 5-9$), compared to $\times1.09$ at $z=3-5$, although the uncertainties overlap. If we instead split the sample in stellar mass, we find a stronger enhancement in sSFR (factor of $\times1.15$) for galaxies with lower stellar masses ($\log_{10}(M_\star / {\rm M_\odot}) = 8.0-9.0$), compared to $\times1.04$ at higher masses, although again the uncertainties overlap. We therefore consider both the redshift and stellar-mass dependencies tentative.

\item We find that the boost in sSFR might also potentially depend on the distance to the second closest companion, with a weak tendency towards higher sSFR ratios at smaller $r_2$, although the measurements are consistent with the controls within their uncertainties. We find no significant dependence on the number of neighbours in the local environment, with only a very weak indication of elevated sSFR ratios at $N_{\rm env}\gtrsim60$.

\item By analysing the shape of the SFH of each individual high-probability close-pair ($r_{\rm p} < 30$~kpc), we find that they can be categorised into two distinct groups: galaxies with monotonically rising SFHs that peaked recently, and objects with rising SFHs that peaked at $> 10$~Myr and started decreasing at recent lookback times. One possible interpretation of this bimodality is that some objects are in the \textit{pre-pericentre} or \textit{pre-coalescence} phase, while the larger fraction may have already undergone a close passage and be observed in a \textit{post-pericentre} phase.

\item Comparing the fraction of AGN in pairs to the pair fraction of the spectroscopic-only sample yields consistent results, meaning that we do not find a signature of significant AGN triggering in our galaxy pair sample. This may be explained in part by the AGN duty cycle being much shorter than the merger timescales we probe ($0.1-1~{\rm Myr} \ll 50-100~{\rm Myr}$). Moreover, we do not probe the final coalescence phase at $r_{\rm p} < 5$~kpc, which is the most relevant event for strong gas channelling to the galactic centre that potentially triggers AGN activity.

\item Similarly, we do not find an excess in the fraction of Ly$\alpha$ emitters in pairs compared to $f_{\rm spec, pair}$ at all separations and redshifts. This may indicate that galaxy interactions are not efficient at Ly$\alpha$ photon production and creating channels for Ly$\alpha$ photon escape or ionized bubbles, at the wider separations ($r_{\rm p}=5-100$~kpc) that we probe. This does not rule out the enhancement of Ly$\alpha$ emission at the smallest scales in very close pairs, compact groups and overdense regions.

\end{itemize}

To study the physical processes ongoing in systems participating in close interactions and mergers in more detail, spatially resolved studies are required. One of the revolutionary tools of \textit{JWST}, the NIRSpec IFU, is the ideal instrument for such resolved studies \citep[e.g.][]{Jones_2025b, Rodriguez_Del_Pino_2024}, and tracing the kinematics of such systems for future large-sample studies.

\section*{Acknowledgements}

We thank David R. Patton for insightful discussions and helpful comments during the `Dancing in the Dark: when Galaxies Shape Galaxies' conference in Sexten, Italy, which helped to improve this work.

DP acknowledges support by the Huo Family Foundation through a P.C. Ho PhD Studentship. DP and ST acknowledge support by the Royal Society Research Grant G125142. CS, GCJ, JS, and RM acknowledge support by the Science and Technology Facilities Council (STFC), by the ERC through Advanced Grant 695671 ``QUENCH'', and by the UKRI Frontier Research grant RISEandFALL. RM also acknowledges funding from a research professorship from the Royal Society. IJ acknowledges support by the Huo Family Foundation through a P.C. Ho PhD Studentship. WMB acknowledges support by a research grant (VIL54869) from VILLUM FONDEN. AJB acknowledges funding from the ``FirstGalaxies'' Advanced Grant from the European Research Council (ERC) under the European Union’s Horizon 2020 research and innovation programme (Grant Agreement No. 789056). SC acknowledges support by the European Union's HE ERC Starting Grant No. 101040227 - WINGS. ECL acknowledges support of an STFC Webb Fellowship (ST/W001438/1). DJE, BDJ, MR, and BER acknowledge support by a JWST/NIRCam contract to the University of Arizona, NAS5-02015. DJE is also supported as a Simons Investigator. BER also acknowledges support from the JWST Program 3215. The research of CCW is supported by NOIRLab, which is managed by the Association of Universities for Research in Astronomy (AURA) under a cooperative agreement with the National Science Foundation. JW gratefully acknowledges support from the Cosmic Dawn Center through the DAWN Fellowship. The Cosmic Dawn Center (DAWN) is funded by the Danish National Research Foundation under grant No. 140.

This work is based [in part] on observations made with the NASA/ESA/CSA James Webb Space Telescope. The data were obtained from the Mikulski Archive for Space Telescopes at the Space Telescope Science Institute, which is operated by the Association of Universities for Research in Astronomy, Inc., under NASA contract NAS 5-03127 for JWST. These observations are associated with PIDs 1180, 1181, 1210, 1286, 1895, 1963, and 3215. The authors acknowledge the teams led by PIs Daniel Eisenstein (PID 3215), Christina Williams, Sandro Tacchella, Michael Maseda (JEMS; PID 1963), and Pascal Oesch (FRESCO; PID 1895) for developing their observing program with a zero-exclusive-access period.

\section*{Data Availability}

The data underlying this article will be shared at a reasonable request by the corresponding author. Fully reduced NIRCam images and NIRSpec spectra are publicly available on MAST (\url{https://archive.stsci.edu/hlsp/jades}), with \doi{10.17909/8tdj-8n28}, \doi{10.17909/z2gw-mk31}, and \doi{10.17909/fsc4-dt61} \citep{Rieke_2023, Eisenstein_2025, Bunker_2024, D'Eugenio_2025}.



\bibliographystyle{mnras}
\bibliography{references.bib} 



\appendix

\section{Stellar mass distribution and completeness}
\label{app:mass_completeness}

\begin{figure}
 \centering
 \includegraphics[width=\linewidth]{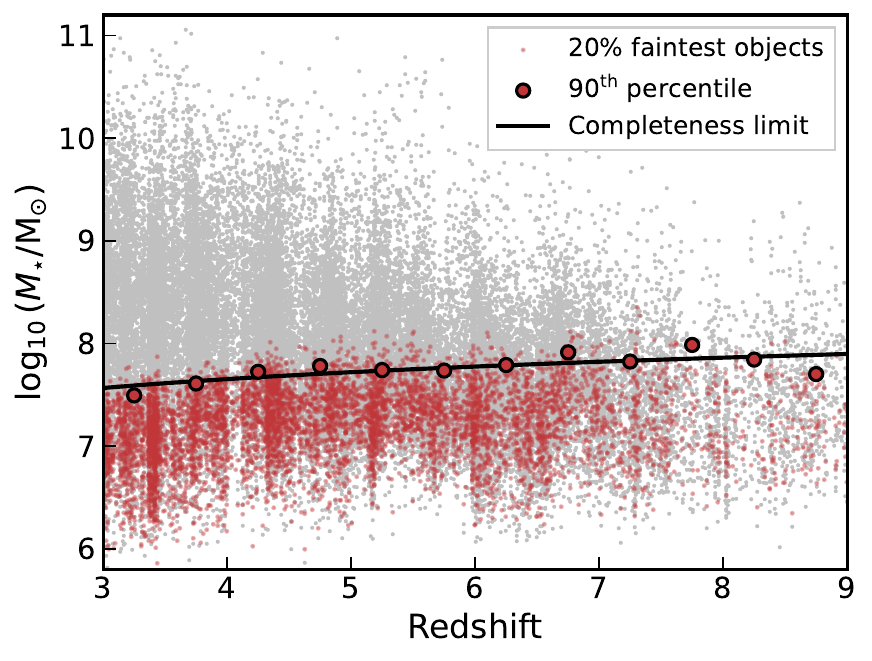}
 \caption{Stellar mass distribution and completeness as a function of redshift of the combined GOODS-South and GOODS-North sample. Grey points show the stellar masses of galaxies as a function of redshift. Coloured points show the limiting stellar masses ($M_{\star}^{\rm lim}$) of the faintest 20 per cent of galaxies in each $\Delta z=0.5$ bin, calculated from the corresponding F444W $5\sigma$ point-source depth following the method of \citet{Pozzetti_2010}. Large circles indicate the $90^{\rm th}$ percentile of the $M_{\star}^{\rm lim}$ distribution in each redshift bin, where we conservatively adopt the higher value obtained from GOODS-South and GOODS-North. The solid black curve shows the best-fitting redshift-dependent stellar mass completeness limit, described by a logarithmic function. The resulting completeness limit remains below $\log_{10}(M_{\star}/{\rm M_{\odot}})=8$ over the full redshift range $3\leq z\leq9$.}
 \label{fig:mass_completeness}
\end{figure}

To assess the stellar mass completeness of the sample, we follow the method of \citet{Pozzetti_2010}, as described in detail in \citet{Puskas_2025}. We calculate the limiting stellar mass ($M_{\star}^{\rm lim}$) of each galaxy using its observed F444W magnitude and the corresponding $5\sigma$ limiting magnitude of the survey tier. In redshift bins of width $\Delta z=0.5$, we select the faintest 20 per cent of galaxies and define the 90 per cent completeness limit as the $90^{\rm th}$ percentile of their $M_{\star}^{\rm lim}$ distribution. We perform this calculation separately for the combined GOODS-South and GOODS-North samples, adopting $5\sigma$ F444W point-source depths of 3.37 and 5.75 nJy, respectively, and conservatively take the higher of the two limits in each redshift bin. The resulting completeness limits are shown in Fig.~\ref{fig:mass_completeness} and are well described by a logarithmic function. The fitted completeness limit increases from $\log_{10}(M_{\star}^{\rm comp}/{\rm M_{\odot}})\simeq7.57$ at $z=3$ to $\log_{10}(M_{\star}^{\rm comp}/{\rm M_{\odot}})\simeq7.90$ at $z=9$. Therefore, throughout the full redshift range considered in this work, the adopted lower stellar mass limit of the primary sample, $\log_{10}(M_{\star}/{\rm M_{\odot}})=8$, remains above the estimated 90 per cent completeness limit. The secondary sample extends to lower stellar masses because of the major merger mass ratio requirement of $\mu\geq0.25$, and we therefore additionally require both members of each selected pair to lie above the corresponding redshift-dependent stellar mass completeness limit.

\section{Additional checks}
\label{sec:additional checks}

\begin{figure*}
\centering
\subfloat{\includegraphics[width=0.5\linewidth]{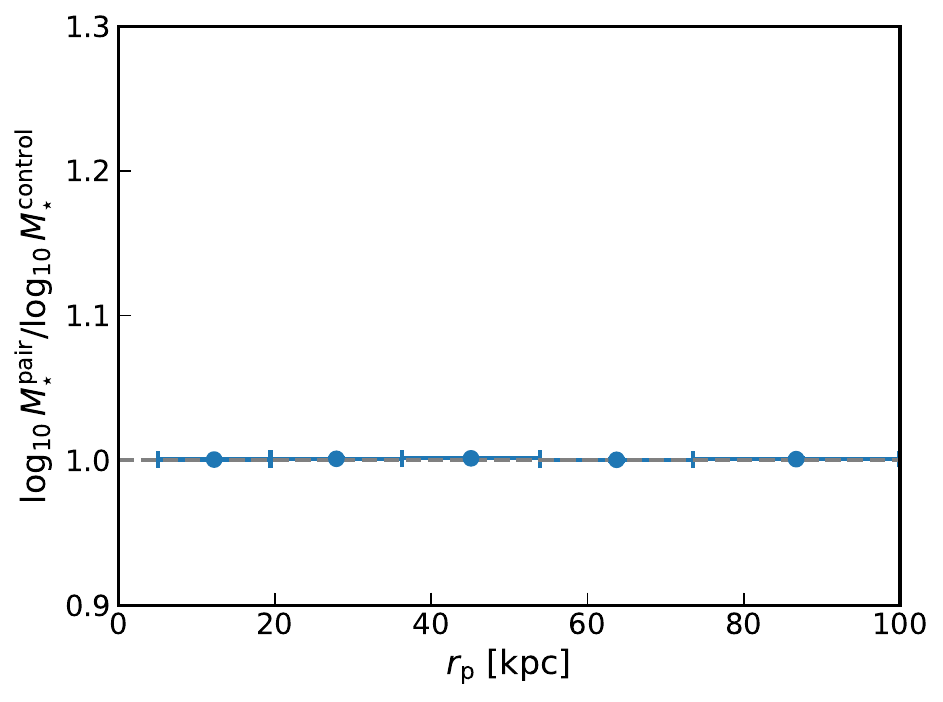}}
\subfloat{\includegraphics[width=0.5\linewidth]{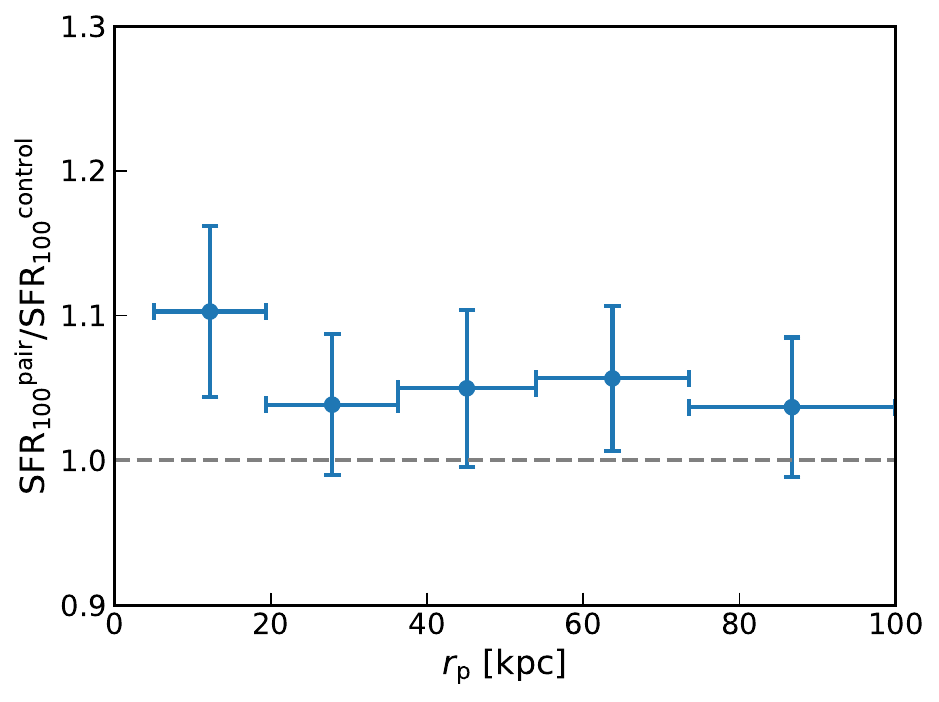}}
\caption{Additional checks to validate our detection of star formation enhancement in paired galaxies versus projected separation to the closest companion. \textit{Left:} ratio of the stellar mass of primary galaxies to the median stellar mass of their matched controls, which is consistent with unity across the full separation range. This is an assuring check that the control matching was adequately performed in the stellar mass space and that any enhancement in SFR or sSFR is not driven by the paired galaxies being more or less massive than their associated controls. \textit{Right:} enhancement in SFR for the 100 Myr averaging timescale, which shows a similar modest enhancement at the smallest separations to that measured in sSFR. Together with the absence of a stellar-mass offset, this supports an enhancement in recent star formation in the closest pairs.}
\label{fig:diagnostics}
\end{figure*}

In Fig.~\ref{fig:diagnostics} we present two additional checks that assess the sSFR enhancement reported in Section~\ref{sec:Enhanced Star Formation in Galaxy Pairs}. First, we verify that the result is not an artefact of imperfect control matching in stellar mass by explicitly comparing the stellar masses of paired galaxies to the medians of their matched controls as a function of projected separation (left panel of Fig.~\ref{fig:diagnostics}). The ratio is consistent with unity across all separations, confirming that our four-dimensional matching procedure (redshift, stellar mass, local density, and isolation) produces closely matched stellar masses and that the observed sSFR offset is not driven by residual differences in stellar mass. Second, we repeat the analysis using ${\rm SFR}_{100}$ instead of ${\rm sSFR}_{100}$ (right panel of Fig.~\ref{fig:diagnostics}). The SFR enhancement closely mirrors the behaviour seen in sSFR while the mass ratio remains flat, with the largest enhancement again occurring at the smallest projected separations. This provides an independent check that the modest sSFR excess originates from elevated recent ($\sim100$~Myr) star formation rather than differences in stellar mass normalisation.

Throughout, median values and their uncertainties are obtained using the same adaptive-binning procedure as for the fiducial sSFR analysis in Section~\ref{sec:Enhanced Star Formation in Galaxy Pairs}. As further discussed in Section~\ref{sec:caveats and parameter choices} and Fig.~\ref{fig:param_var}, varying the probability thresholds used for the pair identification and environmental measurements generally preserves the qualitative indication of elevated sSFR at the smallest separations, although the detailed radial behaviour and strength of the enhancement vary between the different selections.

\bsp	
\label{lastpage}
\end{document}